\pdfoutput=1

\documentclass[a4paper,11pt]{article}

\usepackage{import}

\usepackage[utf8]{inputenc}
\usepackage[top=2cm, bottom=2cm, left=2.5cm, right=2.5cm]{geometry}
\usepackage{hyperref} % for links
\usepackage{enumitem} % for listes
\usepackage{blindtext} % lorem ipsum
\usepackage{comment}
\usepackage{fancyvrb} % verbatim
\usepackage{caption}
\usepackage{subcaption}
\usepackage{float} % position figures
\renewcommand{\,}{\hspace{1pt}}
\renewcommand{\_}{\ul{\hspace{7pt}}}

\usepackage{tikz-cd} % comm diag
\usepackage{forest} % tree structures

\usepackage{amsmath} % matrices
\usepackage{amssymb} % pour mathfrak
\usepackage{stmaryrd} % rrbracket
\usepackage{cmll}

\usepackage{amsthm}
\usepackage[capitalize]{cleveref}
\usepackage{thmtools}
\usepackage{thm-restate}
\declaretheorem[numberwithin=section]{theorem}
\declaretheorem[sibling=theorem]{corollary}
\declaretheorem[sibling=theorem]{proposition}

\declaretheorem[sibling=theorem]{lemma}
\declaretheorem[sibling=theorem, style=definition]{example}
\declaretheorem[sibling=theorem, style=definition]{definition}
\declaretheorem[sibling=theorem, style=definition]{remark}
\declaretheorem[sibling=theorem, style=definition]{related Work}

\usepackage{ebutf8}

\usepackage[style=numeric, maxnames=10]{biblatex}
\bibliography{../Biblio/category_theory}
\bibliography{../Biblio/proof_assistants}
\bibliography{../Biblio/nested_datatypes}
\bibliography{../Biblio/modules_over_monads}
\bibliography{../Biblio/sigma-monoids}
\bibliography{../Biblio/hss}
\bibliography{../Biblio/unclassified}

\newcommand{\mc}[1]{\mathcal{#1}}
\newcommand{\mb}[1]{\mathbb{#1}}
\newcommand{\trm}[1]{\textrm{#1}}
\newcommand{\mrm}[1]{\mathrm{#1}}

\newcommand{\ol}[1]{\overline{#1}}
\newcommand{\ul}[1]{\underline{#1}}
\usepackage{scalerel}
\DeclareMathOperator*{\bigplus}{\scalerel*{+}{\sum}}

\newcommand{\N}{\mathbb{N}}

\newcommand{\Cmon}{(\mc{C}, \otimes, I, \alpha, \lambda, \rho)}
\newcommand{\Set}{\mrm{Set}}
\newcommand{\Mon}{\mrm{Mon}}
\newcommand{\Mod}{\mrm{Mod}}
\newcommand{\Model}{\mrm{Model}}
\newcommand{\Sig}{\mrm{Sig}}
\newcommand{\SigStrength}{\mrm{SigStrength}}
\newcommand{\TotMonMod}[1]{\int_{R : \Mon(#1)} \Mod(R)}
\newcommand{\TotSigModel}[1]{\int_{\Sigma : \Sig(#1)} \Model(\Sigma)}
\newcommand{\op}{\mrm{op}}
\newcommand{\Id}{\mrm{Id}}
\newcommand{\id}{\mrm{id}}

\newcommand{\Free}{\mrm{Free}}
\newcommand{\var}{\mrm{var}}
\newcommand{\app}{\mrm{app}}
\newcommand{\abs}{\mrm{abs}}
\newcommand{\inl}{\mrm{inl}}
\newcommand{\inr}{\mrm{inr}}
\newcommand{\swap}{\mrm{swap}}

\DeclareMathOperator{\colim}{colim}
\title{A Unified Framework for Initial Semantics}
\title{A Unified Framework for Initial Semantics}
\author{Thomas Lamiaux \and Benedikt Ahrens}
\date{}

\begin{document}

\maketitle

\begin{abstract}

  % In this work, we consider initial semantics, that characterizes
  % the syntax of programming languages with variable binding, with its
  % substitution structure, as an initial object in a suitable category.

  % Intro IS / Which IS
  Initial semantics aims to capture inductive structures and their properties as initial
  objects in suitable categories.
  We focus on the initial semantics aiming to model the syntax and substitution
  structure of programming languages with variable binding as initial objects.
  %
  % Three approaches
  Three distinct yet similar approaches to initial semantics have been proposed.
  An initial semantics result was first proved by Fiore, Plotkin, and Turi using
  $Σ$-monoids in their seminal paper published at LICS'99.
  Alternative frameworks were later introduced by Hirschowitz and Maggesi using
  modules over monads, and by Matthes and Uustalu using heterogeneous substitution systems.
  Since then, all approaches have been significantly developed by numerous researchers.
  % Links are unclear
  While similar, the links between this different approaches remain unclear.
  This is especially the case as the literature is difficult to access, since it
  was mostly published in (short) conference papers without proofs, and contains many
  technical variations and evolutions.

  In this work, we introduce a framework based on monoidal categories that
  unifies these three distinct approaches to initial semantics, by suitably
  generalizing and combining them.
  Doing so we show that modules over monoids provide an abstract and easy to
  manipulate framework, that $Σ$-monoids and strengths naturally arise when stating
  and proving an initiality theorem, and that heterogeneous substitution systems
  enable us to prove the initiality theorem modularly.
  Moreover, to clarify the literature, we provide an extensive overview of related work
  using our framework as a cornerstone to explain the links between the
  different approaches and their variations.

\end{abstract}
\newpage

\tableofcontents
\newpage

\section{Introduction}

\emph{Initial semantics} as a concept aims to characterize inductive structures and their
properties as initial objects in suitable categories.
We focus on the strand of initial semantics that aims to characterize (higher-order)
programming languages as initial objects in well-chosen categories of ``models''.
Considering programming languages categorically, as appropriate initial models, has many advantages.
By choosing models of a programming language to be suitable algebras, models
provide us with a mathematical abstraction of the syntax of the language, including
variable-binding constructions.
Indeed, in a model, the terms and the constructors of the language are
represented as abstract categorical notions that capture their properties.
This enables us to manipulate the language by only considering these categorical
notions, without having to constantly deal with its syntax and implementation,
in particular with the implementation of binders.
Furthermore, the initiality of the model offers an abstract recursion principle.
Indeed, due to initiality, there exists a model morphism from the initial model,
which represents syntax, to any model.
Unfolding the definitions, this provides us with a recursion principle in the
same categorical terms as the ones used for the abstract syntax,
thus enabling us to reason by recursion on the abstract syntax without
having to deal with any underlying implementation.
Initial models can therefore abstract the syntax with its recursion
principle.

Using more advanced notions of model, it is possible to account for syntax with
its substitution structures \cite{FPT99,HirschowitzMaggesi07,Hss04}.
This involves abstracting the syntax, but also abstracting how simultaneous
substitution works, its properties, and how it interacts with constructors.
In particular, we refer to initial semantics as the theory of designing models
of languages, accounting for syntax and substitution, and proving initiality theorems.
That is, we aim to prove theorems asserting, ideally under simple conditions, that a language has an initial model.
Initiality theorems enable us to prove that a language has a well-defined
substitution structure, but by initiality, they also enable us to equip the
language with a substitution-safe recursion principle.
By this, we mean a recursion principle that respects the substitution structure by
construction, which is important if one wishes to preserve the operational semantics of
languages.

\subsection{Different Approaches to Initial Semantics}

Three distinct yet similar approaches to initial semantics have been proposed.

The most well-known approach to initial semantics, but also the first to be invented,
was introduced by Fiore, Plotkin, and Turi in their seminal paper \cite{FPT99}.
In \cite{FPT99}, focusing on untyped languages, the authors introduced
$Σ$-monoids to capture languages with their substitution, and suggested
a very general method based on monoidal categories and functors with strength
to prove initiality theorems.
This framework has been developed further to encompass richer type systems
\cite{Cbn02,MMCCS05,SecondOrderDep08,Polymorphism11,ListObjects17},
but also to add meta-variables and equations
\cite{HamanaMetavar04,SecondOrderDep08,HurPhd,PolymorphismEq13,FioreSzamozvancevPopl22}.
It has also recently been extended to skew-monoidal categories \cite{CellularHoweTheorem20}.

Another approach to initial semantics was later developed by André Hirschowitz
and Maggesi \cite{HirschowitzMaggesi07}, using modules over monads.
This approach has the advantage of providing a very general and abstract definition
of signatures and models \cite{HirschowitzMaggesi12,ZsidoPhd10,PresentableSignatures21},
that can be easier to manipulate.
For instance, it has been extended to account for equations
\cite{PresentableSignatures21,2Signatures19} and reduction rules
\cite{UntypedRelativeMonads16,TypedRelativeMonads19,ReductionMonads20,TransitionMonads22}.
However, modules over monads and associated notions only have been defined and
studied for endofunctor categories, that is, for particular monoidal categories.
Moreover, this general notion of signature lacks some desirable theoretical properties like a general initiality theorem.
Only restricted and instance-specific ones have been proven \cite{HirschowitzMaggesi10,ZsidoPhd10},
for instance for the category $[\Set,\Set]$.

A third, less-known approach based on heterogeneous substitution systems,
abbreviated as hss, has been developed.
Hss were originally invented to prove that both wellfounded and
non-wellfounded syntax \cite{Hss04} have a monadic substitution structure.
It was applied in this way in the UniMath library, to construct untyped and
simply-typed higher-order languages \cite{HssUntypedUniMath19,HssTypedUnimath22}.
Yet, hss were also shown to provide a framework for initial semantics by
Ahrens and Matthes in \cite{HssRevisited15}.
This approach is based on functors with strength and endofunctor categories,
and has general initiality theorems, and seems to have a stronger notion
of models.

These approaches have been extended to model various added structures like meta-variables
\cite{SecondOrderDep08}, equations between terms \cite{FioreHur10,2Signatures19},
or reductions rules \cite{UntypedRelativeMonads16,ReductionMonads20}.
However, we do not discuss them in this work and only mention them in the discussion of related work.

Importantly, we discuss in this paper the ``nested datatype'' approach to variable binding.
Here, a lambda abstraction is represented by a constructor $\lambda : T(X+1) \to T(X)$, where $X$ and $X+1$ are contexts of free variables.
There are other ways to represent variables and binding:
firstly, the ``higher-order abstract syntax'' approach --- where lambda abstraction is represented by a constructor $\lambda : (T \to T) \to T$ --- was explored, in particular, by Hofmann \cite{DBLP:conf/lics/Hofmann99}.
Secondly, the nominal approach --- where lambda abstraction is represented by a constructor $\lambda : [\mathbb{A}]T \to T$, for a set of ``atoms'' $\mathbb{A}$ --- was explored, in particular, by Gabbay and Pitts \cite{DBLP:conf/lics/GabbayP99}.
The discussion of these alternative approaches to variable binding and their relationship to the nested approach are outside the scope of this work.

\subsection{The Challenges of Relating the Different Approaches}

The starting point for this work is the observation that, while similar, the formal
links between this three approaches are understudied, and still largely unknown.

While similar, the different traditions present several important structural differences.
The different frameworks are not all defined for the same underlying structures.
The $Σ$-monoid tradition is defined for generic monoidal categories
\cite{SecondOrderDep08}, whereas the hss and modules over monads traditions
only deal with endofunctor categories \cite{Hss04,PresentableSignatures21},
very particular monoidal categories.
More importantly, the different frameworks uses different notions of signatures
and models, and it is not always clear how these notions relate to each other.
Even in one specific tradition, different works may employ different notions of signatures and models.
For instance, the notions of signature differ between \cite{HirschowitzMaggesi07,ZsidoPhd10}
and \cite{PresentableSignatures21,2Signatures19}.
The different frameworks also have different relationship towards the initiality theorem.
The $Σ$-monoid tradition provides a generic adjoint theorem \cite{SecondOrderDep08},
from which an initiality theorem is deduced, whereas the hss one only has an
initiality theorem \cite{HssRevisited15}, and the modules over monads tradition
only has initiality theorems for very specific signatures and endofunctor categories like
$[\Set,\Set]$ or $[\Set^T,\Set^T]$ \cite{HirschowitzMaggesi10},\cite[Section 6]{ZsidoPhd10}.
Whatever the tradition, there also are different versions of the initiality theorem.
For instance, some rely on monoidal closedness \cite{SecondOrderDep08,HssRevisited15},
while others rely on $\omega$-cocontinuity \cite{ListObjects17,HssUntypedUniMath19}.

This is complicated by the fact that the existing literature on initial semantics
seems, to us, difficult to access for newcomers to the field and even experts.
Many important notions are spread out over different papers, often with small
technical yet non trivial variations.
Moreover, many papers of the subject have been published
as extended abstracts or in conference proceedings without appendices.
As such, they usually contain very limited discussion of related work, few fully worked-out
examples, and often no proofs or only sketches of proofs.

The relationship between the different approaches has been little studied in the literature.
When it has, the result are little known, do not cover the full extent of the
variations mentioned before, and are sometimes very specific.
For instance, for untyped algebraic signatures, the links between $Σ$-monoids on $[\mb{F},\Set]$,
and module based models on $[\Set,\Set]$, have been investigated in Zsidó's
dissertation \cite{ZsidoPhd10}.
Furthermore, Zsidó also investigated the links for simply-typed algebraic signatures, for
the categories $[\mb{F} \downarrow T,\Set]^T$ and $[\Set^T,\Set^T]$.
As another example, for the particular monoidal category $[\Set,\Set]$, signatures with strength
and $Σ$-monoids have been shown to be a particular case of signatures and
modules based models \cite{HirschowitzMaggesi12}.
Finally, recently, and independently from us, heterogeneous substitution systems
have been shown to be an adequate abstraction to prove an initiality result
for $Σ$-monoids, as very briefly mentioned in \cite[Section 4.4]{HssNonWellfounded24}.

\subsection{Contributions}

In this work, we bridge this gap in the literature by presenting a framework
that unifies the different traditions of initial semantics.
We achieve this by appropriately generalizing and integrating different
components of the existing frameworks.
Doing so, we demonstrate that each tradition of initial semantics addresses a
distinct aspect of the problem, assembling into a comprehensive framework.
More specifically:

\begin{enumerate}
  \setlength\itemsep{-1pt}
  \item We base the framework on monoidal categories as these encompass the
        great majority of the existing instances, including categories like
        $[\mb{F},\Set]$ that are not endofunctor categories.

  \item We present a general and abstract framework for signatures and models,
        using modules over monoids, a variant of modules over monads for
        monoidal categories.
        We further port and generalize the different results from the tradition of modules
        over monads to our framework.

  \item This framework is too general to prove general initiality and adjoint theorems.
        Consequently, to state and prove such theorems we restrict ourselves
        to signatures with strength and $Σ$-monoids, particular signatures and models.
        We further show how they appear naturally when one tries to state and prove an initiality result.

  \item We leverage heterogeneous substitution systems that we generalize to monoidal
        categories to prove the initiality theorem, efficiently and modularly.
        To do so, we generalize and adapt to our purpose a proof scattered
        throughout several papers \cite{Hss04,DeBruijnasNestedDatatype99,HssRevisited15,HssUntypedUniMath19}.

  \item We prove the adjoint theorem from the initiality theorem rather than the
        opposite, and rely on $\omega$-cocontinuity rather than on monoidal closedness.
        This is a key difference for relating the different approaches.

\end{enumerate}

To fully justify our design choices and our claim that we encompass the different
instances, we provide an extensive discussion of related work in which we
compare the different approaches and their variations to our presentation.
In particular, this enables us to shed a new light onto the literature, clarifying
the links between the different approaches through comparaison with our framework.

Altogether, with this work, we aim to provide a self-contained presentation
that can serve as a foundation for exploring more advanced concepts in initial
semantics, such as those involving meta-variables or reduction rules.
We leave the exploration of these advanced topics as an open problem.

\subsection{Synopsis}
\label{sec:synopsis}

In \cref{sec:prelims}, we briefly recall the definition and a few results about
monoidal categories and $ω$-colimits that are used in the paper.
We present our framework for initial semantics in
\cref{sec:models,sec:initiality_theorem,sec:building_initial_model}.
Specifically, we present an abstract framework based signatures and models based
on modules over monoids in \cref{sec:models}.
As not all signatures have an initial model, to prove an initiality theorem and
an adjoint theorem, we restrict ourselves to signatures with strength.
We show they naturally arise as particular signatures when stating and
proving these theorems, in \cref{sec:initiality_theorem}.
We then provide a self-contained and complete proof of the initiality theorem
and adjoint theorem, based on heterogeneous substitution systems, in
\cref{sec:building_initial_model}.
We provide an extensive analysis of the literature, and use our framework as a
cornerstone to relate the different approaches to each other, in \cref{sec:related-work}.
We conclude and discuss open problems in \cref{sec:conclusion}.

\section{Preliminaries}
\label{sec:prelims}

The framework defined herein is entirely written in the language of category theory.
We assume the reader is familiar with the notions of category, functor, and
natural transformation, as found, e.g., in Riehl's book \cite{CategoryTheoryInContext14}.
We recall here some more specific definitions and properties about monoidal
categories and $ω$-colimits that are needed to understand the framework.

\subsection{Monoidal Categories}

To be able to define our notion of models, we need more structure on the
underlying category than the one provided by the mere definition of a
category.
Hence, throughout the paper, we work with monoidal categories.
A modern presentation of monoidal categories can be found in
\cite[Section 3.1]{2DimensionalCategories20}.

\begin{definition}[Monoidal Categories]
  \label{def:mon-cat}
  A \emph{monoidal category} is a tuple $\Cmon$, where $\mc{C}$ is a category,
  $\_ ⊗ \_ : \mc{C} × \mc{C} \to C$ is a bifunctor called the
  monoidal product, and $I : \mc{C}$ an object called the unit.
  $α,λ, ρ$ are natural isomorphisms -- called the associator, and the
  left and right unitor -- that satisfy the unit axiom and the pentagon axiom.
  \begin{align*}
    α_{X,Y,Z} : (X ⊗ Y) ⊗ Z ≅ X ⊗ (Y ⊗ Z)
    &&
    λ_{X} : I ⊗ X ⊗ X
    &&
    ρ_{X} : X ⊗ I ⊗ X
  \end{align*}
\end{definition}

\begin{remark}
  The above notations are fairly standard, however be careful that in some
  references, as in \cite{HssRevisited15,HssUntypedUniMath19}, $ρ_X$ and
  $λ_X$ are swapped.
\end{remark}

\begin{example}[Category of endofunctors]
  Given any category $\mc{C}$, the category of endofunctors $[\mc{C},\mc{C}]$
  is monoidal for the composition of functors as monoidal product, the
  identity functor $\Id$ as unit, and the identity natural transformation
  for $α,λ,ρ$.
\end{example}

\subsection{$ω$-Colimits}
\label{subsec:omega-colimits}

The notion of $ω$-colimit is important for the construction of syntax, since it
abstractly formalizes the idea of building sets of abstract syntax trees by
recursion on the height of such trees.
We thus use $ω$-colimits in the construction of initial models, and we crucially
use that signatures preserve $ω$-colimits when we construct an initial model, as
discussed in \cref{sec:initiality_theorem}.

\begin{definition}[$ω$-chains]
  An $ω$-chains is a sequence $(C_i, c_i)_{i : \N}$ of objects and
  morphisms assembling as a left oriented infinite chain:
  \[
    \begin{tikzcd}
      C_0 \ar[r, "c_0"]
        & C_1 \ar[r, "c_1"]
        & C_2 \ar[r, "c_2"]
        & C_3 \ar[r, "c_3"]
        & \; ...
    \end{tikzcd}
  \]
\end{definition}

\begin{example}
  Given a category $\mc{C}$ with initial object $0 : \mc{C}$, for any
  endofunctor $F : \mc{C} \to \mc{C}$  on $\mc{C}$, there is a canonical
  $ω$-chain associated to $F$ denoted by $\mrm{chn}_F$:
  \[
    \begin{tikzcd}
      0 \ar[r, "\star"]
        & F(0) \ar[r,   "F(\star)"]
        & F^2(0) \ar[r, "F^2(\star)"]
        & F^3(0) \ar[r, "F^3(\star)"]
        & \; ...
    \end{tikzcd}
  \]
\end{example}

\begin{definition}[$ω$-cocontinuous functors]
  \label{def:omega-cocontinuous}
  The colimit of an $ω$-chain is called an $ω$-colimit.
  A functor $F : \mc{C} \to \mc{D}$ is \emph{$ω$-cocontinuous} if it
  preserves all $ω$-colimits.
  The collection of $ω$-cocontinuous functors $\mc{C} → \mc{C}$ forms a full subcategory
  of the category of endofunctors $\mc{C} → \mc{C}$.
\end{definition}

\begin{example}
  The identity functor $\Id : \mc{C} → \mc{C}$ is $ω$-cocontinuous.
\end{example}

\subsubsection*{Closure properties}

We aim to build our signatures modularly, i.e. out of smaller signatures.
Being $ω$-continuous is a key feature of our signatures; we crucially use that
$ω$-continuous functors are closed under some operations, such as taking
colimits, limits and composition.
These facts can be proved using different results in
\cite[Section 3.8]{CategoryTheoryInContext14}.

\begin{proposition}[Closure under colimits]
  \label{prop:omega-colimits}
  If $\mc{C}$ is cocomplete, then the category of $ω$-cocontinuous
  functors $\mc{C} → \mc{C}$ is closed under colimits.
\end{proposition}

\noindent While $ω$-continuous functors are closed under all small colimits for
a wide variety of base categories, they are usually only closed under finite
limits. For our purpose, this will suffice.

\begin{proposition}[Closure under limits]
  \label{prop:omega-limits}
  If $\mc{C}$ admits a class of limits that commutes with $ω$-colimits
  in $\mc{C}$, then the category of $ω$-cocontinuous functors
  $\mc{C} → \mc{C}$ is closed under this particular class of limits.
\end{proposition}

\begin{proposition}
  \label{prop:presheaves-limits}
  In the category $\Set$, and in functor categories $[\mc{C},\Set$], finite limits commute
  with $ω$-colimits.
\end{proposition}

\section{Mathematical Structure of Syntax}
\label{sec:models}

One aspect of denotational semantics is to find (or design) a mathematical structure
that captures the structure and property of syntax appropriately.
The operations we aim to capture are variable binding and substitution, and the properties
are those that describe substitution laws.
In other words, we aim to define a general notion of model of a language, encompassing the existing ones, for
untyped and typed higher-order languages, encompassing their syntaxes and
substitution structures.

For substitution, the notion of monad has been found to capture its structure and properties
\cite{BellegardeHook94,DeBruijnasNestedDatatype99,AltenkirchReus99}.
For instance, the untyped lambda calculus can be described as a monad on the category of sets.
Yet, not all categories suitable for modeling syntax are endofunctor categories,
e.g., the functor $[\mb{F},\Set]$, with $\mb{F}$ the category of \emph{finite} sets (or a skeleton thereof) can be used to model the untyped lambda calculus \cite{FPT99}.
Some categories are not even functor categories as discussed in \cite{ISPoly24}.
However, all the categories used for modeling syntax are \emph{monoidal} categories, and the
substitution structure is given by the respective monoidal product.
The inclusion of variables into terms is modeled by the unit of the monoid.
Consequently, to build a general theory of syntax, we work directly with
monoids in a monoidal category.
We can recover different examples of syntax, and their substitution structure,
as instantiations of the theory to a specific monoidal category.

To model languages, we additionally need a mathematical structure to model
constructors, and a notion of signature to specify them.
Two main structures have been considered to model constructors: firstly,
strength \cite{FPT99,Cbn02,ListObjects17}, and secondly, modules over monads
\cite{HirschowitzMaggesi10,ZsidoPhd10,PresentableSignatures21}.
Moreover, different notions of signature have been considered: either
syntactic or mathematical, either using strengths or modules over monads.
To encompass the different point of views, we rely on of modules over monads and
associated notions, that we fully generalize to monoidal categories.
Indeed, once generalized, modules over monads provide us with a fully abstract
framework and notions of signatures; the other notions naturally appearing
as particularly important instances.

In the following, we first review monoids and modules over monoids in
\cref{subsec:monoids,subsec:modules} used to model simultaneous substitution
and constructors.
This enables us to define our category of signatures in \cref{subsec:signatures}.
Afterwards, given a signature, we define and describe its associated category of
models in \cref{subsec:models}.
Throughout, we rely on the untyped lambda calculus as a running example.

\subsection{Monoids}
\label{subsec:monoids}

\subsubsection{Limitations of Monads}
\label{subsubsec:kleisli}

Before introducing monoids to model substitution, let us first consider why
Monads are not sufficient to encompass the different existing approaches.
Monads on $\mc{C}$ have long been identified as a suitable axiomatization of
simultaneous substitution and its basic properties
\cite{BellegardeHook94,DeBruijnasNestedDatatype99,AltenkirchReus99}.

\begin{definition}[Monads]
  \label{def:monads}
  A \emph{monad} on $\mc{C}$ is a triple $(T,η,σ)$ where
  $T : \mrm{ob}(\mc{C}) → \mrm{ob}(\mc{C})$ is a function on objects,
  $η$ is a family of morphisms $η_{\,Γ} : Γ → T(Γ)$, and such that for all
  morphisms $c : Γ → T(Δ)$ there is a morphisms $σ_{\,Γ,Δ}(c) : T(Γ) → T(Δ)$,
  satisfying the equations:
  \begin{align*}
    \begin{tikzcd}[ampersand replacement=\&]
      Γ \ar[r, "η_{\,Γ}"] \ar[dr, swap, "c"]
        \& R(Γ) \ar[d, "σ_{\,Γ,Δ}(c)"] \\
        \& R(Δ)
    \end{tikzcd}
    &&
    \begin{tikzcd}[ampersand replacement=\&]
      R(Γ) \ar[dr, bend left, "σ_{\,Γ,Γ}(η_{\,Γ})"]
                \ar[dr, swap, bend right, "\id"]
        \& \\
        \& R(Γ)
    \end{tikzcd}
    &&
    \begin{tikzcd}[ampersand replacement=\&]
      R(Γ) \ar[r, "σ_{\,Γ,Δ}(c)"]
                \ar[dr, swap, "σ_{\,Γ,Υ}(σ_{Δ,Υ}(c') \circ c)"]
        \& R(Δ) \ar[d, "σ_{Δ,Υ}(c')"] \\
        \& R(Υ)
    \end{tikzcd}
  \end{align*}
\end{definition}

\noindent Monads enable us to model substitution with $T$ associating contexts
to well-defined terms in them, $η$ corresponding to variables, and $σ$ to
simultaneous substitution.
The equations then specify that substituting a variable returns the associated
terms, that substituting variables by variables is the identity, and the
composition rule of simultaneous substitution.

The issue with this approach is that monads are not general enough to encompass
all the approaches, as they force us to represent our terms as an endofunctor
$\mc{C} → \mc{C}$.
This does not account for model based on $[\mb{F},\Set]$ \cite{FPT99} to model
untyped languages, nor on one based $ω$-cocontinuous endofunctors
$[\mc{C},\mc{C}]_ω$ that can both be interesting.
Such categories require variants of monads to account for the
substitution structure like relative monads \cite{RelativeMonads15} for
$[\mb{F},\Set]$.
Moreover, some languages like system F are more naturally modeled by a coherent
family of endofunctors and monads, rather than by one functor and one monad
\cite{ISPoly24}.

Thankfully, the substitution structure of all this examples are actually
equivalent to monoids in their associated monoidal categories.
Consequently, to be able to account for all possibilities, we rely directly on the
category of monoids in a monoidal category \cite[Section 1.2]{2DimensionalCategories20}
to model the substitution structure of languages.
The advantage is that it enables us to be fully abstract, and so to build a generic framework.
However, the downside is that the framework does not directly model substitution.
It is only once appropriately instantiated that the framework models substitution.

\subsubsection{Monoids}

\begin{definition}[Monoids]
  \label{def:monoids}
  Given a monoidal category $\Cmon$, a \emph{monoid} on $\mc{C}$ is a tuple
  $(R,μ,η)$ where $R$ is an object of $\mc{C}$ and the \emph{multiplication}
  $μ : R ⊗ R → R$ and the \emph{unit} $η : I → R$ are morphisms of $\mc{C}$ such
  that the following diagrams commute:
  \begin{align*}
    \begin{tikzcd}[ampersand replacement=\&]
      (R ⊗ R) ⊗ R \ar[r, "\alpha"] \ar[d, swap, "μ ⊗ R"]
        \& R ⊗ (R ⊗ R) \ar[r, "R ⊗ μ"]
        \& R ⊗ R \ar[d, "μ"] \\
      R ⊗ R \ar[rr, swap, "μ"]
        \&
        \& R
    \end{tikzcd}
  &&
    \begin{tikzcd}[ampersand replacement=\&]
      I ⊗ R \ar[r, "η ⊗ R"] \ar[dr, swap, "λ_R"]
        \& R ⊗ R \ar[d, "μ"]
        \& R ⊗ I \ar[l, swap, "R ⊗ η"] \ar[dl, "ρ_R"] \\
        \& R
        \&
    \end{tikzcd}
  \end{align*}
\end{definition}

\begin{definition}[Morphisms of Monoids]
  \label{def:morphisms-monoids}
  A \emph{morphism of monoids} $(R,μ,η) → (R',μ',η')$ is a morphism $f : R → R'$
  in $\mc{C}$ preserving multiplication and the unit i.e such that the following
  diagrams commute:
  \begin{align*}
    \begin{tikzcd}[ampersand replacement=\&]
      R ⊗ R \ar[r, "f  ⊗ f"] \ar[d, swap, "μ"]
        \& R' ⊗ R' \ar[d, "μ'"] \\
      R \ar[r, swap, "f"]
        \& R'
    \end{tikzcd}
    &&
    \begin{tikzcd}[ampersand replacement=\&]
      \& I \ar[dl, swap, "η"] \ar[dr, "η'"]
      \& \\
      R \ar[rr, swap, "f"]
      \&
      \& R'
    \end{tikzcd}
  \end{align*}
\end{definition}

\begin{proposition}[Category of Monoids]
  \label{prop:cat-monoids}
  Given a monoidal category $\mc{C}$, monoids in $\mc{C}$ and their morphisms
  form a category denoted $\Mon(\mc{C})$.
\end{proposition}

To understand how monoids enables us to model substitution, let us consider
the untyped lambda calculus on $[\Set,\Set]$ as an example.

\begin{example}[Untyped Lambda Calculus]
  \label{ex:LC-monoids}
  In the category $[\Set,\Set]$, the untyped lambda calculus forms a functor
  $Λ : \Set → \Set$ associating context to well-defined terms over them.
  For $c : Γ → Δ$, the functorial action $Λ(c) : Λ(Γ) → Λ(Δ)$ corresponds to
  renaming, and the functorial law that renaming is well-behaved: renaming by the
  identity is the identity $Λ(\id) = \id$, and renaming twice is the same as
  renaming once $Λ(c' ∘ c) = Λ(c') ∘ Λ(c)$.

  It further forms a monoid with $η$ the variable constructor $\var$, and
  $μ : Λ (Λ (Γ)) → Λ (Γ)$ the flattening operation that given a term with terms
  as label replaces the variables by their labels.
  As morphisms of $[\Set,\Set]$, $\var$ and $μ$ are natural transformations.
  This specifies how renaming computes on them: renaming a variable is renaming the label $Λ(c)(\var(x)) = \var (c(x))$,
  and renaming the flattening is renaming before flattening $Λ(c) (\mu(u)) = \mu (Λ (Λ (c)) (u))$.
  The monoid laws then specify the compatibility of flattening with renaming
  $μ_Γ ∘ μ_{Λ(Γ)} = μ_Γ ∘ Λ (μ_Γ)$ and $μ (Λ (\var)(t)) = t$, and that flattening
  variables returns the labels $μ (\var (t)) = t$.

  It finally enables us to model simultaneous substitution as monoids are
  equivalent to Kleisli triples.
  $T$ and $η$ corresponds to $Λ$ and $\var$, and substitution is defined
  as $\sigma(c) := R(Γ) \xrightarrow{R(c)} R(R(Δ)) \xrightarrow{μ_Δ} R(Δ)$.
  Basically, it replaces all labels by their substitutions, then flatten to
  removes the wrapping variable constructor.
\end{example}

\begin{example}[Simply-typed lambda calculus]
  To model the simply-typed lambda calculus with type system $T$, one could
  consider the base category $[\Set^T,\Set^T]$ \cite{Cbn02,ZsidoPhd10}.
\end{example}

\subsection{Modules over Monoids}
\label{subsec:modules}

Monoids in $\mc{C}$ enable us to model variables and simultaneous substitution.
It remains to model the language-specific constructors, such as application and abstraction of the lambda calculus;
in particular, how substitution computes on these constructors.
Unfortunately, constructors can not be modeled by morphisms of monoids: the
latter must preserve the $\eta$ constructor of monoids corresponding to
variables, which is not the case for constructors like application or
abstraction of the untyped lambda calculus.

Therefore, we rely instead on modules over a monoid \cite[Chapter4]{ZsidoPhd10},
a more liberal variant of monoids, and characterize constructors as morphisms of
suitable modules.
Roughly, modules over monoids are similar to monoids, but without a unit
constructor $η$ that morphisms would have to respect, giving us additional freedom to
express constructors as module morphisms.

\begin{definition}[Module over a Monoid]
  \label{def:modules}
  Given a monoid $R : \Mon(\mc{C})$, a (left) $R$-module is a tuple $(M,
  p^M)$ where $M$ is an object of $\mc{C}$ and $p^M : M ⊗ R → M$ is
  a morphism of $\mc{C}$ called \emph{module substitution} that is compatible
  with the multiplication and the unit of the monoid:
  \begin{align*}
    \begin{tikzcd}[ampersand replacement=\&]
      (M ⊗ R) ⊗ R \ar[r, "\alpha_{M,R,R}"] \ar[d, swap, "p^M ⊗ R"]
        \& M ⊗ (R ⊗ R) \ar[r, "M ⊗ μ"]
        \& M ⊗ R \ar[d, "p^M"] \\
      M ⊗ R \ar[rr, swap, "p^M"]
        \&
        \& M
    \end{tikzcd}
    &&
    \begin{tikzcd}[ampersand replacement=\&]
      M ⊗ I \ar[r, "M ⊗ η"] \ar[dr, swap, "\rho_M"]
        \& M ⊗ R \ar[d, "p^M"] \\
        \& M
        \&
    \end{tikzcd}
  \end{align*}
\end{definition}

\begin{related Work}
  Using morphisms of modules over monads to model constructors was first
  considered in \cite{HirschowitzMaggesi07,HirschowitzMaggesi10} for $[\Set,\Set]$.
  We rely instead on modules over monoids \cite[Chapter4]{ZsidoPhd10} as they
  are defined for any monoidal category, and not just for endofunctor categories
  $[\mc{C},\mc{C}]$.
  See \cref{subsec:rw-modules-over-monoids} for a discussion on the differences.
\end{related Work}

\begin{definition}[Morphisms of Modules]
  \label{def:morphisms-modules}
  Given a monoid $R : \Mon(\mc{C})$, a \emph{morphism of $R$-modules} $(M, p^M) →
  (M', p^{M'})$ is a morphism $r : M → M'$ of $\mc{C}$ commuting with
  the respective module substitutions:
  \[
    \begin{tikzcd}
      M ⊗ R \ar[r, "r ⊗ R"] \ar[d, swap, "p^M"]
        & M' ⊗ R \ar[d, "p^{M'}"] \\
      M \ar[r, swap, "r"] & M'
    \end{tikzcd}
  \]
\end{definition}

\begin{proposition}[Category of Modules]
  \label{prop:cat-modules}
  Given a monoid $R : \Mon(\mc{C})$, the modules over $R$ and their
  morphisms form a category denoted $\Mod(R)$.
\end{proposition}

By construction, module morphisms respect module substitutions.
Given some language modeled by a monoid $R$, we can model its language constructors
as $R$-module morphisms $M → M'$;
here, the $R$-modules $M$ and $M'$
specify how substitution behaves at the input and output of term constructors,
and hence which variables are bound.
The dependency on the monoid $R$ enables us to access the language substitution
structure, and to build modules using it.
And indeed, the monoid forms a trivial module over itself $Θ$ corresponding to
the usual substitution on terms.

\begin{definition}
  Given a monoid $(R, \eta, \mu)$, the pair $(R,μ)$ is a module over the monoid $R$.
  By abuse of notation, we also denote this by $R$.
\end{definition}

\begin{definition}
  Given an object $D : \mc{C}$ and a $R$-module $(M,p^M)$, there is a
  $R$-module with object $D ⊗ M : \mc{C}$ and module
  substitution $(D ⊗ M) ⊗ R \xrightarrow{α} D ⊗ (M ⊗ R) \xrightarrow{D ⊗ p^M} D ⊗ M$.
\end{definition}

\noindent In this work, will restrict to constructors that returns terms without
introducing fresh variables as it simplifies the rest of the framework and is
enough in practice.
Given a language modeled by some monoid $R$, we will hence model constructors of
that language by a module morphisms $M → R$, for a suitable $R$-module $M$.

To understand how it works, let us consider the untyped lambda calculus on $[\Set,\Set]$.
First, as variable binding depends on the type system of the languages and hence the base category,
we must build a specific module to model untyped variable binding.
We can then represent the constructors:

\begin{definition}
  \label{ex:module_binding}
  Let $R$ be a monad on $\Set$, that is, a monoid in $[\Set,\Set]$.
  There is a module $R^{(n)}$ modeling the binding of $n$ variables
  with underlying functor $R^{(n)}(Γ) := Γ + n$ and module substitution:
  \[ (R^{(n)} \circ R) (Γ) := R (R (Γ) + n)
      \xrightarrow{\;\; R[R(\inl_Γ), η_{\,Γ+n}] \;\;} R (R (Γ + n))
      \xrightarrow{\;\;μ_{Γ+n}\;\;} R (Γ + n)
  \]
\end{definition}

\begin{remark}
  Given a monoid $R$, The module $R^{(0)}$ is simply the module $R$.
\end{remark}

\begin{example}[Untyped Lambda Calculus, continuation of \cref{ex:LC-monoids}]
  \label{ex:LC-modules}
  On $[\Set,\Set]$, the constructor $\app$ can be modeled by a module morphism
  $Θ × Θ → Θ$ as it take two arguments without binding variables, and $\abs$ by
  a module morphism $Θ^{(1)} → Θ$ as it takes one argument binding one variable.

  As morphism, $\app$ and $\abs$ are natural transformations. This specifies how
  renaming computes on constructors: renaming $\app$ is renaming its arguments
  $Λ(c)(\app(u,v)) = \app(Λ(c)(u), Λ(c)(v))$, and renaming $\abs$ is renaming
  its argument while preserving the fresh variable $Λ(c)(\abs(u)) =
  \abs(Λ(c+1)(u))$.
  Similarly, the morphism law specify how flattening computes on constructors.
  Flattening $\app$ is flattening its arguments $μ (\app(u,v)) = \app (μ(u),μ(v))$,
  and flattening $\abs$ is flattening its argument while preserving the fresh variable
  $μ(\abs(u)) = \abs\, (μ (Λ[Λ(\inl), \var])(u))$.

  As discussed in \cref{subsec:monoids,ex:LC-monoids}, monoids do not directly
  model substitution, they only do so when appropriately instantiated.
  Similarly, modules morphisms do not directly model the computation rules of
  substitution on constructors, they only do so when appropriately instantiated.
  Actually, for a general morphism of module, there is no reason why
  computation rules for flattening induce computational rules for substitution.
  Fortunately, in practice this is always the case.
  For the untyped lambda calculus, given $c : Γ → Λ(Δ)$, substituting
  $\app$ is substituting its arguments $σ(c)(\app(u,v)) = \app(σ(c)(u), σ(c)(v))$,
  and substituting $\abs$ is substituting the argument $σ(c)(\abs(u)) = \abs(σ(\ol{c})(u))$
  for the weakening of $c$, $\ol{c} = [Λ(\inl_{Δ,1}) ∘ c, η_{Δ+1} ∘ \inr_{Δ,1}] : X+1 \to  Λ (Δ + 1)$.
\end{example}

To specify constructors as module morphisms $M → Θ$, we want to
build modules \emph{modularly}, such as $Θ × Θ$ as the domain of the application of lambda calculus.
The construction of some modules depends on the underlying monoidal category $\mc{C}$ and can not be built generically; for instance, the module
$Θ^{(n)}$, since variable binding depends on the type system.
However, many others can be
built as limits and colimits:

\begin{proposition}[Closure under (co)-limits]
  \label{prop:modules-closure-colimits}
  Given a monoid $R : \Mon(\mc{C})$, the category $\Mod(R)$ inherits its limits
  and colimits from $\mc{C}$ provided that $\_ ⊗ I$, $\_ ⊗ R$, $\_ ⊗ R ⊗ R$ and
  $\_ ⊗ (R ⊗ R)$ preserve them.
\end{proposition}
\begin{proof}
  Let $F : J \to \Mod(R)$ be a diagram in $\Mod(R)$, i.e. a natural family $(M_i,p^{M_i})$.
  \begin{itemize}[label=$-$]
    \setlength\itemsep{-1pt}
    \item Forgetting the module substitution gives us a diagram in $\mc{C}$, that by
          assumption has a colimit $\colim M_i : \mc{C}$.
    \item As it is preserved by $\_ ⊗ R$ to build a module substitution $p^{\colim M_i} : \colim
          M_i ⊗ R → \colim M_i$, it suffices to build a morphism $\colim (M_i ⊗ R) → \colim M_i$.
          This follows pointwise by functoriality of colimits.
    \item This morphism is then compatible with the multiplication and unit of $R$,
          as the $\colim M_i$ are preserved and the properties are satisfied pointwise.
  \end{itemize}
  Consequently, $(\colim M_i, p^{\colim M_i})$ forms a module.
  It further is the colimit as $\colim M_i$ is in $\mc{C}$, and $\_ ⊗ R$ preserves it.
   For limits, the construction is analogous.
\end{proof}

\begin{example}
  \label{prop:module-language-cst}
  Under the above assumptions, $\Mod(R)$ has a terminal module denoted $R^0$,
  and is closed under products and coproducts.
\end{example}

\noindent These last constructions are of importance for modularity.
The terminal module enables us to represent constants, i.e. constructors without
arguments, as module morphisms $R^0 → R$.
The closure under products enables us to represent term constructors with
several independent inputs, like $\app$ as a morphism $Θ × Θ → Θ$.
Closure under coproducts allow us to specify languages with several independent
term constructors, as a morphism $(R^0 + R × R) → R$ amounts by
universal property to a morphism $R^0 → Θ$ and a morphism $R × R → R$

\begin{example}
  On $[\Set,\Set]$, the constructors of the untyped lambda calculus $\Lambda$ can be modeled by a module morphism
  $(\Lambda × \Lambda + \Lambda^{(1)}) → \Lambda$.
\end{example}

\subsection{Signatures}
\label{subsec:signatures}

A model of a language should be a pair of a monoid $R$ axiomatizing variables
and simultaneous substitution, and an $R$-module morphism $M → R$ axiomatizing
term constructors.
Here, the $R$-module $M$ should suitably model the domain of the constructors of the language.
Hence a signature $\Sigma$ should associate to any monoid $R$, an $R$-module
$\Sigma(R) : \Mod(R)$ representing the input of the constructors.
To express this dependency, we introduce the total category of modules
\cite{HirschowitzMaggesi12,PresentableSignatures21} in \cref{subsubsec:total-cat-modules},
before defining signatures in \cref{subsubsec:sig}.

\subsubsection{The Total Category of Modules}
\label{subsubsec:total-cat-modules}

\begin{proposition}[The pullback functor]\label{prop:pullback_module}
  Given a morphism of monoids $f : R → R'$, there is a functor $f^* : \Mod(R') → \Mod(R)$
  that associates to each module $(M',p^{M'}) : \Mod(R')$ a $R$-module $f^*M'$
  called the \emph{the pullback module} defined as
  \[
    \begin{tikzcd}
      M' ⊗ R \ar[r, "M' ⊗ f"]
        & M' ⊗ R' \ar[r,"p^{M'}"]
        & M'.
    \end{tikzcd}
  \]
\end{proposition}

\begin{proposition}[The module functor]\label{prop:module_functor}
  There is a contravariant pseudofunctor of bicategories, from the discrete
  bicategory of monoids in $\mc{C}$ to the bicategory of (small) categories,
  \[ \Mod : \Mon(\mc{C})^\op ⟶ \mrm{Cat}. \]
  This pseudofunctor associates, to a monoid $R$, the category
  $\Mod(R)$, and to a morphism of monoids $f : R → R'$, the functor
  $f^* : \Mod(R)' → \Mod(R)$ of \cref{prop:pullback_module}
\end{proposition}

Applying the Grothendieck construction to the pseudofunctor of \cref{prop:module_functor},
we obtain a fibration over the category of monoids on $\mc{C}$:

\begin{proposition}[The total category of modules]
  \label{prop:cat-totalcat-modules}
  There is a \emph{total category of modules} $\TotMonMod{\mc{C}}$.
  Its objects are tuples $(R,M)$ where $R : \Mon(\mc{C})$ is a monoid, and
  $M : \Mod(R)$ a module over it.
  Its morphisms $(R,M) → (R',M')$ are tuples $(f,r)$ where $f : R → R'$
  is a morphism of monoids and $r : M → f^*M'$ a morphism of $R$-monoids.

  The forgetful functor from the total category of modules to the base
  category of monoids is a Grothendieck fibration,
  \begin{equation}\label{eq:forget-module-monoid}
    U : \TotMonMod{\mc{C}} ⟶ \Mon(\mc{C}).
  \end{equation}
\end{proposition}

\subsubsection{Signatures}
\label{subsubsec:sig}

We can now define a signature to be a functor $\Sigma : \Mon(\mc{C}) ⟶ \TotMonMod{\mc{C}}$
that returns a module over the input one, specifying the arity of the constructors.
,

\begin{definition}[Signatures]
  \label{def:sig}
  A \emph{signature} is a functor $\Sigma : \Mon(\mc{C}) ⟶ \TotMonMod{\mc{C}}$
  making the following diagram commute:
  \[
    \begin{tikzcd}
      \Mon(\mc{C}) \ar[rr, "\Sigma"] \ar[dr, swap, equal]
        &
        & \TotMonMod{\mc{C}} \ar[dl, "U"] \\
      & \Mon(\mc{C}) &
    \end{tikzcd}
  \]
\end{definition}

\begin{related Work}
  This definition was introduced in \cite{HirschowitzMaggesi12,PresentableSignatures21} for $[\Set,\Set]$,
  and is sometimes referred to as ``parametric modules'' \cite{TransitionMonads22}.
  In the context of modules, other notions of signatures have been considered.
  We refer to \cref{subsubsec:rw-modules-sig} for a discussion on the subject.
\end{related Work}

\noindent In other words, signatures are sections of the forgetful functor $U : \TotMonMod{\mc{C}} ⟶ \Mon(\mc{C})$.
For a given signature $\Sigma : R \mapsto (R,M)$, since the first component is
always the identity, we usually omit it and denote by $\Sigma(R)$ the $R$-module $M$.

\begin{definition}[Morphism of Signatures]
  A \emph{morphism of signature} $\Sigma → \Sigma'$ is a natural transformation
  $h : \Sigma → \Sigma'$ that is the identity when composed with the
  forgetful functor $U : \int_{R : \Mon(\mc{C})} \Mod(R) ⟶ \Mon(\mc{C})$
\end{definition}

\begin{proposition}[Category of Signatures]
  \label{prop:cat-signatures}
  Signatures and their morphisms form a category $\mrm{Sig}(\mc{C})$.
\end{proposition}

The advantage of this definition is that constructions on modules extend pointwise.
This enables us to preserve the intuition of modules over monoids, and to recover
our basic building blocks for signatures:

\begin{definition}
  There is a trivial signature $Θ : R ↦ R$.
\end{definition}

\begin{definition}
  For all $D : \mc{C}$ and $\Sigma : \Sig(\mc{C})$, there is a signature $D ⊗ \Sigma : R ↦ D ⊗ Σ(R)$.
\end{definition}

\begin{definition}
  On $[\Set,\Set]$, there is a signature $Θ^{(n)} : R ↦ R^{(n)}$.
\end{definition}

\begin{example}[Untyped Lambda Calculus, continuation of \cref{ex:LC-modules}]
  On $[\Set,\Set]$, the constructors $\app$ and $\abs$ can be modeled by the signature $Θ × Θ$ and $Θ^{(1)}$.
\end{example}

Moreover, limits and colimits are inherited pointwise from modules enabling
us to build signatures as modularly as modules, and to define languages modularly:

\begin{proposition}[Evaluation]
  \label{prop:evaluation-sig}
  Given a monoid $R : \Mon(\mc{C})$, there is an evaluation functor
  $\Sig(\mc{C}) → \Mod(R)$, defined on object by $Σ ↦ Σ(R)$, and on natural
  transformation $f : Σ → Σ$ by the module morphism $Σ(f) : Σ(R) → Σ(R')$.
\end{proposition}

\begin{proposition}[Closure under (co)-limits]
  \label{prop:sig-closure-colimits}
  The category $\Sig(C)$ inherits its limits and colimits from $\mc{C}$ provided
  that for all $R : \Mon(\mc{C})$, $\_ ⊗ R$ preserves them.
\end{proposition}
\begin{proof}
  Given a diagram $Σ : J → \Sig(\mc{C})$ of signatures, we define a new signature $\colim Σ$ as follows.
  \begin{itemize}[label=$-$]
    \setlength\itemsep{-1pt}
    \item On objects, we associate to a monoid $R$ the colimit of modules $R ↦ \colim_{\Mod(R)}\, Σ_i(R)$
          which exists by \cref{prop:evaluation-sig} and \cref{prop:modules-closure-colimits}.
    \item On morphisms, given a morphism of monoids $f : R → R'$,
          we need to build a morphism $\colim_{\Mod(R)}\, Σ_i(R) → f^*  (\colim_{\Mod(R')}\, Σ_i(R'))$.
          This amounts to a morphism in $\mc{C}$ from $\colim_{\mc{C}}\, Σ_i(R) →
          \colim_{\mc{C}}\, Σ_i(R')$ that respects module substitution.
          This follows pointwise from functoriality of colimits.
  \end{itemize}
  Consequently, $\colim Σ$ is a well-defined functor, and is, by construction, a
  signature, since it is the identity on its first component.
  It further is the colimit since it is the colimit pointwise.
  Limits are constructed analogously.
\end{proof}

\begin{example}
  Under the above assumptions, $\Sig(\mc{C})$ has a terminal signature, also
  denoted $Θ^0$, and is closed under products and coproducts.
\end{example}

In practice, we work with a \emph{fixed} category satisfying this properties, like $[\Set,\Set]$.
In which case, as variable binding is instance specific, we often rely on a specific
subclass of signatures, that is closed by some limits and colimits to be modular.
For instance, on $[\Set,\Set]$, one often rely on algebraic signatures that
are simple but enough to specify usual languages:

\begin{definition}[Algebraic Signatures]
  \label{def:alg-sig}
  \emph{Algebraic signatures} on $[\Set,\Set]$, also known as \emph{binding
  signatures}, are generated by $Θ^{(n)}$, coproducts and \emph{finite} products.
  In other words, there are of the form $\scalebox{1.5}{+}_{i ∈ I}\;\, Θ^{(n_0)} × ... × Θ^{(n_k)}$.
\end{definition}

\begin{remark}
  Though, $[\Set,\Set]$ is closed by generic products, they need to be finite
  for algebraic signatures to have a model.
  See \cref{subsec:initiality_theorem}, in particular \cref{prop:omega-limits}.
\end{remark}

\begin{example}
  \label{ex:alg-sig}
  On $[\Set,\Set]$, the following languages can represented by algebraic signatures:
  \begin{itemize}[label=$-$]
    \setlength\itemsep{-1pt}
    \item The untyped lambda calculus with constructors $\app,\abs$ by $Θ × Θ + Θ^{(1)}$.
    \item First order logic with constructors $⊤,⊥,¬,\land,\lor,⇒,∃,∀$ by $2Θ^0 + Θ + 3Θ^2 + 2Θ^{(1)}$.
    \item Linear logic with constructors $⊤,⊥,!,?,\&,\parr,⊕,⊸,∃,∀$ by $2Θ^0 + 2Θ +5Θ^2 + 2Θ^{(1)}$.
  \end{itemize}
\end{example}

\subsection{Models}
\label{subsec:models}

\subsubsection{Models of a signature}

Given a signature, we can now represent languages as a monoid representing
the language with substitution and a module morphism representing the
constructors.

\begin{definition}[Models]
  \label{def:models}
  Given a signature $\Sigma : \Sig(\mc{C})$, a \emph{model} of $\Sigma$ is a
  tuple $(R,r)$ where $R : \Mon(\mc{C})$ is a monoid and $r : \Sigma(R) →
  R$ is a morphism of $R$-modules.
\end{definition}

\begin{related Work}
  A notion of model based on modules was introduced in \cite{HirschowitzMaggesi07}, and in its
  current form in \cite{HirschowitzMaggesi12,PresentableSignatures21}, in both
  cases for $[\Set,\Set]$.
  There are different variations depending on the underlying notion of signature
  used. See \cref{subsubsec:rw-modules-sig} for details on the variations.
\end{related Work}

Morphisms of models must then respect variables and the module morphism in source and target.
That is, it must be a morphism of monoids that respects the diagram below which
can be understood as a diagram in $\mc{C}$ as the forgetful functor $U : \Mod(R) → \mc{C}$ is faithful.

\begin{definition}[Morphism of Models]
  \label{def:morphims-models}
  A morphism of $\Sigma$-models $(R,r) → (R',r')$ is a morphism of monoids
  $f : R → R'$ compatible with the module morphism $r$ and $r'$, i.e. such
  that the left-hand side diagram of $R$-modules commutes:
  \begin{align*}
    \begin{tikzcd}[ampersand replacement=\&]
      \Sigma(R) \ar[r, "r"] \ar[d, swap, "\Sigma(f)"]
        \& R \ar[d, "f"] \\
      f^* \Sigma(R') \ar[r, swap, "f^*r'"]
        \& f^*R'
    \end{tikzcd}
    &&
    \begin{tikzcd}[ampersand replacement=\&]
      (R,\Sigma(R)) \ar[r, "(id{,}r)"] \ar[d, swap, "\Sigma(f)"]
        \& (R,R) \ar[d, "Θ(f)"] \\
      (R', \Sigma(R')) \ar[r, "(id{,}r')"]
        \& (R',R')
    \end{tikzcd}
  \end{align*}
\end{definition}

\begin{remark}
  The pullback along $f^*$ is here for homogeneity as $\Sigma(R')$ is an $R'$-module.
  It can be understood by viewing the left-hand side diagram as a diagram in
  the total category of modules, as on the right-hand side.
\end{remark}

\begin{proposition}[Category of Models]
  \label{prop:cat-mod-sig}
  Given a signature $\Sigma : \Sig(\mc{C})$, its models and their morphisms
  form a category denoted $\Model(\Sigma)$.
\end{proposition}

\subsubsection{Representable Signatures}
\label{subsubsec:rep-sig-rec}

We aim to not just equip syntax with a mathematical structure, but also to show
that it has a universal property, being the \emph{initial} such mathematical
structure, in a suitable category of ``models''.
Initiality then provides a recursion principle, following the classical method
for simple datatypes\cite{Goguen76,GoguenEtAl75}.

Not all signatures do admit initial models; those who do, we call \emph{representable}.

\begin{definition}[Representable Signatures]
  A signature $\Sigma$ is \emph{representable} if its category of models
  $\Model(\Sigma)$ has an initial model.
  In this case, its initial model is denoted $\ol{\Sigma}$, and called
  the \emph{syntax} associated to $\Sigma$.
\end{definition}

\begin{example}
  \label{ex:alg-sig-rep}
  Algebraic signatures on $[\Set,\Set]$ (see \cref{def:alg-sig}) are representable.
  In particular, the signatures of untyped lambda calculus, first order logic
  and linear logic are representable (\cref{ex:alg-sig}).
\end{example}

The universal property provided by initiality is stronger than the usual
recursion principle of inductive types, as it provides us not only with a
$\mc{C}$ morphism but with a morphism of models.
Such a morphism is automatically compatible, in a suitable sense, with substitution in source and target.
As an example, let us consider a substitution-safe translation from first-order
logic to linear logic.

\begin{example}[{{\cite[Example 9.1]{PresentableSignatures21}}}]
  \label{ex:FOL-to-LL}
  By \cref{ex:alg-sig-rep}, first-order logic and linear logic both have an initial model.
  Let us denote by $r_x$ the module morphism associated to the logic symbol $x$ in
  the respective initial model.
  It is possible to equip the monoid underlying the model of linear logic with a
  first-order logic structure defining the module morphisms as follows.

  \[ \begin{array}{cccc}
    r_⊤' := r_⊤ & r_⊥' := r_⊥ & r_¬' := r_⊸ ∘ (r_! × r_0) & r_\land' := r_\& \\
    r_\lor' := r_⊕ ∘ (r_! × r_!) & r_⇒' := r_⊸ ∘ (r_! × \Id) & r_∃' := r_∃ ∘ r_! &
    r_∀' := r_∀
  \end{array} \]

  \noindent By initiality of the model of first order logic, there is then a
  substitution-safe translation from first-order logic to linear logic.
\end{example}

As shown in \cite[Section 5.1]{PresentableSignatures21} for $[\Set,\Set]$, any
initial model satisfies a fixpoint property.
We generalize this property to monoidal categories.
In particular, it enables us to prove that some signatures are not representable.

\begin{lemma}[Lambek's Theorem]
  \label{lemma:abstract-Lambek}
  Let $F : \mc{D} → \mc{D}$ be a functor and $α : F → \Id$ a natural transformation
   such that $Fα = αF$. If $0 : \mc{D}$ is an initial object, then $0 ≅ F(0)$.
\end{lemma}

\begin{proposition}
  \label{prop:fixpoint-models}
  Let $\mc{C}$ be a monoidal category with binary coproducts, and such
  that for all $Z : \mc{C}$, $\_ ⊗ Z$ preserves coproducts.
  If a signature $\Sigma$ has a model $M$, then $I + \Sigma(M)$ can be
  turned into a model of $\Sigma$.
  Moreover, if $\Sigma$ is representable, there is an isomorphism of models
  $I + \Sigma(\ol{\Sigma}) \cong \ol{\Sigma}$.
\end{proposition}
\begin{proof}
  There is a monoid structure on $I + Σ(M)$ with unit $I \xrightarrow{\inr} I + Σ(M)$, and
  using that precomposition distributes over coproducts, multiplication can be defined as
  \[
  \begin{tikzcd}[column sep=3.5cm]
    (I + Σ(M)) ⊗ (I + Σ(M)) \ar[r] \ar[d, phantom, "≅"]
      & I + Σ(M) \\
    I ⊗ (I + Σ(M)) + Σ(M) ⊗ (I + Σ(M)) \ar[r, "λ \;+\; Σ(M) ⊗ {[}η{,}r{]}"]
      &  I + Σ(M) + Σ(M) ⊗ M \ar[u, swap, "{[}\id {,} \inr \,∘\, p^M{]}" ]
  \end{tikzcd}
  \]
  Moreover, there is a module morphism making $I + Σ(M)$ a model of $Σ$:
  \[
    \begin{tikzcd}[column sep=large]
      Σ(I + Σ(M)) \ar[r, "Σ({[}η{,}r{]})"]
        & Σ(M) \ar[r, "\inr"]
        & I + Σ(M)
    \end{tikzcd}
    \]
  This construction assembles into a functor $I + Σ(\_) : \Model(Σ) → \Model(Σ)$,
  and $[η,r] : I + Σ(M) → M$ into a natural transformation $I + Σ(\_) \to \Id$,
  such that $[η,r] (I + Σ(\_)) = (I + Σ(\_)) [η,r]$.
  Consequently, by \cref{lemma:abstract-Lambek}, if $Σ$ has an initial model
  $\ol{Σ}$ then $I + \Sigma(\ol{\Sigma}) \cong \ol{\Sigma}$.
\end{proof}

\begin{example}[{{\cite[Non-example 5.5]{PresentableSignatures21}}}]
  \label{ex:not-representable}
  In the monoidal category $[\Set,\Set]$, the signature $\mc{P} \circ
  Θ$ is not representable.
  Indeed, otherwise there would be an isomorphism of models
  $\Id + (\mc{P} \circ Θ) (\ol{\mc{P} \circ Θ}) \cong \ol{\mc{P} \circ Θ}$.
  Yet, as there is a forgetful functor $\Model([\Set,\Set]) → [\Set,\Set]$, for
  all $X : \Set$, there would be an isomorphism of sets
  $X + \mc{P} ((\ol{\mc{P} \circ Θ})(X)) X \cong (\ol{\mc{P} \circ Θ}) (X)$,
  which is impossible for cardinality reasons.
\end{example}

\subsubsection{The Total Category of Models}
\label{subsubsec:total-model}

The advantage of this framework is that it is fully abstract, and hence
is suitable to express and prove abstract results.
For instance, as introduced in \cite{HirschowitzMaggesi12,PresentableSignatures21}
on $[\Set,\Set]$ models like modules assemble in a total category, which enables us
to easily prove modularity results in the style of \cite[Section 5.2]{PresentableSignatures21}:

\begin{proposition}[The pullback functor]\label{prop:pullback_model}
  Given a morphism of signatures $h : \Sigma → \Sigma'$, there is a functor
  $h^* : \Model(\Sigma') → \Model(\Sigma)$ that associates to to each model
  $(R',r') : \Model(\Sigma')$ a $\Sigma$-model $h^*(R',r')$ called the
  \emph{the pullback model} defined as:
  \[
    \begin{tikzcd}
      \Sigma(R') \ar[r, "h_R"]
        & \Sigma'(R') \ar[r, "r'"]
        & R'
    \end{tikzcd}
  \]
\end{proposition}

\begin{proposition}[The model pseudofunctor]\label{prop:model_pseudofunctor}
  There is a contravariant pseudofunctor of bicategories, from the discrete bicategory of signatures over $\mc{C}$
  to the bicategory of (small) categories,
  \[ \Model : \Sig(\mc{C})^\op ⟶ \mrm{Cat}. \]
  This pseudofunctor associates, to a signature $\Sigma :
  \Sig(\mc{C})$, the category $\Model(\Sigma)$, and to a morphism of
  signatures $h : \Sigma → \Sigma'$, the functor $h^* : \Model(\Sigma') →
  \Model(\Sigma)$ of \cref{prop:pullback_model}.
\end{proposition}

Applying the Grothendieck construction to the pseudofunctor of \cref{prop:model_pseudofunctor}, we obtain a fibration over the category of signatures on $\mc{C}$:

\begin{proposition}[The total category of models]
  \label{prop:cat-totalcat-models}
  Signatures assemble as a \emph{total category of models} $\TotSigModel{\mc{C}}$.
  Its objects are tuples $(\Sigma,(R,r))$ where $\Sigma : \Sig(\mc{C})$ is a
  signature, and $(R,r) : \Model(\Sigma)$ a model over it.
  Its morphisms $(\Sigma,(R,r)) → (\Sigma',(R',r'))$ are tuples $(h,f)$ where $h
  : \Sigma → \Sigma'$ is a morphism of signatures and $f : (R,r) → f^*(R',r')$ a
  morphism of $\Sigma$-models.

  The forgetful functor from the total category of models to the base
  category of signatures is a Grothendieck fibration,
  \[ \TotSigModel{\mc{C}} ⟶ \Sig(\mc{C}) \]
\end{proposition}

Using the total category of models, it is then possible to express a modularity
result on signatures and models.
Specifically, we prove that the initial model of a pushout of signatures is the pushout of
the initial models, generalizing \cite[Section 5.2]{PresentableSignatures21}:

\begin{proposition}[Modularity]
  \label{prop:modularity-models}
  Given a pushout of representable signatures $\Sigma$,$\Sigma_1$,$\Sigma_2$,
  $\Sigma_{12}$, their initial models form a pushout above it, in the total
  category of models:
  \begin{align*}
    \begin{tikzcd}[ampersand replacement=\&]
      \Sigma \ar[r] \ar[d] \arrow[dr, phantom, "\ulcorner", very near end]
        \& \Sigma_2 \ar[d] \\
      \Sigma_1 \ar[r]
        \& \Sigma_{12}
    \end{tikzcd}
    &&
    \begin{tikzcd}[ampersand replacement=\&]
      (\Sigma, \ol{\Sigma}) \ar[r] \ar[d] \arrow[dr, phantom, "\ulcorner",
                                                very near end]
        \& (\Sigma_2, \ol{\Sigma_2}) \ar[d] \\
      (\Sigma_1, \ol{\Sigma_1}) \ar[r]
        \& (\Sigma_{12}, \ol{\Sigma_{12}})
    \end{tikzcd}
  \end{align*}
\end{proposition}
\begin{proof}
  The pushout of signatures provides us with morphisms for the fist component.
  Morphisms for the second component exist by initiality of the respective models.
  The pushout property then follows as it is a pushout on the first component, and
  morphisms on the second component are equal by initiality.
\end{proof}

\section{The Initiality Theorem}
\label{sec:initiality_theorem}

The general notion of signature is built to be fully abstract and practical to
use, e.g. to prove a \hyperref[prop:modularity-models]{modularity result}, or
add equations \cite{2Signatures19} or reduction rules \cite{ReductionMonads20}.
However, it lacks many desirable properties.
There is no known criterion for a signature to have, or not to have, an initial model.
Moreover, there is no known criterion for the product or coproduct of two
representable signatures to be representable.

Consequently, to provide an initiality theorem we must restrict ourselves to
better behaved signatures.
As seen in \cref{prop:fixpoint-models}, the initial model of a representable
signature $Σ$ is a fixpoint of models for the functor $\ul{\Id} + Σ$.
Since we are looking for an initial model with an underlying algebra structure,
Adámek's theorem (\cref{thm:adamek}) and Lambek's theorem (\cref{lemma:abstract-Lambek})
strongly suggests to consider $ω$-cocontinuous functors $Σ : \mc{C} → \mc{C}$.
To turn $Σ$ into a signature, for any monoid $R : \Mon(\mc{C})$ we must equip
$Σ(R)$ with an $R$-module structure, that is, with a module substitution $Σ(R) ⊗ R → Σ(R)$.
Thanks to the monoid structure and functoriality of $Σ$, we have a morphism  $Σ(μ) : Σ (R ⊗ R) → Σ(R)$ in $\mc{C}$.
It hence suffices to have a morphism $θ : Σ(R) ⊗ R → Σ(R ⊗ R)$ such that the composition
\[
  \begin{tikzcd}
    Σ(R) ⊗ R \ar[r, "θ"]
    &
    Σ(R ⊗ R) \ar[r, "Σ(μ)"]
    &
    Σ(R)
  \end{tikzcd}
\] satisfies the module substitution laws.
Unfolding the constraints leads us to \emph{signatures with strength},
which intuitively are better behaved as
their substitution is of a particularly constrained form.
We will prove an initiality theorem for such signatures with strength

\begin{related Work}
  Generalizing work on modules over monads to monoidal categories
  now naturally leads us to consider signatures with strength.
  However, historically, signatures with strength were actually considered \emph{before}
  modules over monads, in the seminal work of Fiore, Plotkin, and Turi \cite{FPT99}.
  Signatures with strength have since been studied extensively \cite{SecondOrderDep08,FioreMahmoud10,HurPhd,ListObjects17}.
  See \cref{subsec:rw-overview} for an historical perspective on initial semantics..
\end{related Work}

In the remainder of this section, we start by reviewing signatures with
strength in \cref{subsec:sig_with_strength}, before discussing the links
between signatures with strength and $Σ$-monoids with signatures and models
in \cref{subsec:sigstrength-to-sig}.
We then state an initiality theorem and an adjoint theorem for signatures
with strength in \cref{subsec:initiality_theorem}, that we prove in \cref{sec:building_initial_model}.

\subsection{Signatures with Strength}
\label{subsec:sig_with_strength}

For a notion of strengths $H(A) ⊗ B → H(A ⊗ B)$ to be suitable to model constructors,
we need at the minimum a strength for variable binding for untyped languages.
As discussed in \cref{ex:sigstrength-variable-binding}, to define such a signature requires $B$
to be more than just an object of $\mc{C}$, it needs to be a \emph{pointed} object.

\begin{definition}[Pointed Object]
  \label{def:pointed-object}
  In a monoidal category $\mc{C}$, a \emph{pointed object} is a tuple $(Z,e)$
  where $Z : \mc{C}$ and $e : I → Z$.
  A morphism of pointed objects $(Z,e) → (Z',e')$ is a morphism $f : Z → Z'$
  such that $f \circ e = e'$.
  Pointed objects form a category $\mathrm{Ptd}(\mc{C})$.
\end{definition}

Having defined pointed objects, we can now define signatures with strength:

\begin{definition}[Signatures with Strength]
  \label{def:sig-strength}
  A \emph{signature with (pointed) strength} in a monoidal category $\Cmon$ is a
  pair $(H,θ)$ where $H : \mc{C} → \mc{C}$ is an endofunctor with a \emph{strength} $θ$,
  that is for all $A : \mc{C}$ and pointed object $b : I → B$ a natural transformation :
  \[ θ_{A,b} : H(A) ⊗ B \longrightarrow H(A ⊗ B) \]
  such that for all $A, b : I → B, c : I → C$, $θ$ is compatible with the
  monoidal associativity and unit, with $b ⊗ c$ as an abbreviation for
  $I \xrightarrow{λ^{-1}_I} I ⊗ I \xrightarrow{b ⊗ c} B ⊗ C$:
  \begin{align*}
    \begin{tikzcd}[ampersand replacement=\&, column sep=large]
      (H(A) ⊗ B) ⊗ C \ar[d, swap, "θ_{A,b} ⊗ C"]
        \& H(A) ⊗ (B ⊗ C) \ar[l, swap, "\alpha^{-1}_{(H(A),B,C)}"]
                                     \ar[dd, "θ_{A, b ⊗ c}"] \\
      H(A ⊗ B) ⊗ C \ar[d, swap, "θ_{A ⊗ B, c}"]
        \& \\
      H((A ⊗ B) ⊗ C) \ar[r, swap, "H(\alpha_{A,B,C})"]
        \& H(A ⊗ (B ⊗ C))
    \end{tikzcd}
    &&
    \begin{tikzcd}[ampersand replacement=\&]
      H(A) ⊗ I \ar[r, "θ_{A,id}"] \ar[d, swap, "\rho_{H(A)}"]
        \& H(A ⊗ I) \\
      H(A) \ar[ur, swap, "H(\rho^{-1}_A)"]
    \end{tikzcd}
  \end{align*}
\end{definition}

\begin{related Work}
  Using strength to model substitution was introduced in \cite{FPT99}, and
  as a formal notion of signatures in \cite{Hss04}.
  We follow a similar presentation to \cite{Hss04}; but a more general one
  defined in terms of actegories can also be found in \cite{SecondOrderDep08}
  and \cite{HssNonWellfounded24}.
\end{related Work}

\begin{definition}[Morphism of signatures with strength]
  A morphism of signatures with strength $(H, θ) → (H', θ')$ is a natural
  transformation $h : H → H'$ compatible with the strengths, i.e., such that
  for all $A$ and $b : I → B$:
  \[
    \begin{tikzcd}
      H(A) ⊗ B \ar[r, "h_A ⊗ B"] \ar[d, swap, "θ_{A,b}"]
        & H'(A) ⊗ B \ar[d, "θ'_{A,b}"] \\
      H(A ⊗ B) \ar[r, swap, "h'_{A ⊗ B}"]
        & H'(A ⊗ B)
    \end{tikzcd}
  \]
\end{definition}

\begin{proposition}[Category of signatures with strength]
  Signatures with strength and their morphisms form a category denoted
  $\SigStrength(\mc{C})$.
  Composition and identity in this category are inherited from the category
  of functors and natural transformations.
\end{proposition}

Signatures with strength admit constructions analogous to those on signatures,
allowing us to specify languages modularly using signatures with strength:

\begin{example}
  There is a trivial signature with strength $(\Id,\Id)$ denoted $Θ$.
\end{example}

\begin{example}
  \label{ex:left-comp}
  Given an object $D : \mc{C}$, for any signature with strength
  $(H,θ)$, there is an associated signature with strength, for the
  endofunctor $D ⊗ H(\_) : \mc{C} → \mc{C}$ and the strength:
  \[
    \begin{tikzcd}
      (D ⊗ H(A)) ⊗ B \ar[r, "\alpha"]
        & D ⊗ (H(A) ⊗ B) \ar[r, "D ⊗ θ"]
        & D ⊗ H(A ⊗ B)
    \end{tikzcd}
  \]
\end{example}

\begin{example}
  \label{ex:sigstrength-variable-binding}
  On $[\Set,\Set]$, there is a signature with strength modeling $n$ variable binding
  defined by the functor $(H)(X)(Γ) := X(Γ + n)$ and the strength
  \[ θ^{(n)} : H(X)(Y(Γ)) := X (Y(Γ) + n) \xrightarrow{X(Y(\inl_Γ) + η_{\,Γ+n})} X (Y (Γ + n)) := H(X ∘ Y)(Γ) \]
\end{example}

Moreover, as for signatures, to be modular we require signatures with
strength to be closed under some limits and colimits.
This follows under the same requirement as for signatures:

\begin{proposition}[Closure under (co)-limits]
  \label{prop:sigstrength-closure-colimits}
  The category $\SigStrength(\mc{C})$ inherits its limits and colimits from $\mc{C}$
  provided that for all $B : \mc{C}$, $\_ ⊗ R$ preserves them.
\end{proposition}
\begin{proof}
  Given a diagram $J → \SigStrength(C)$, so in particular a functorial family $(H_i,θ_i)_{i : J}$,
  we define a new signature pointwise using, firstly, that $\_ ⊗ R$ preserves colimits
  and, secondly, the universal properties of colimits:
  $H : A ↦ \colim H_i(A)$, and $θ : \colim H_i(A) ⊗ B ≅ \colim (H_i(A) ⊗ B) → \colim (H_i(A ⊗ B))$.
  Limits are constructed analogously.
\end{proof}

\begin{example}
  Under the above assumptions, $\SigStrength(\mc{C})$ has a terminal signature, also
  denoted $Θ^0$, and is closed under products and coproducts.
\end{example}

\begin{definition}
  Algebraic signatures on $[\Set,\Set]$ are of the form
  $\scalebox{1.5}{+}_{i ∈ I}\;\, Θ^{(n_0)} × ... × Θ^{(n_k)}$.
\end{definition}

\begin{example}
  On $[\Set,\Set]$, the untyped lambda calculus can be specified by the
  signature $Θ × Θ + Θ^{(1)}$, first order logic by $2Θ^0 + Θ + 3Θ^2 + 2Θ^{(1)}$,
  and linear logic by $\mrm{LL} := 2Θ^0 + 2Θ +5Θ^2 + 2Θ^{(1)}$.
\end{example}

\subsection{Signatures With Strength as Signatures}
\label{subsec:sigstrength-to-sig}

We would like to use signatures with strength instead of signatures to specify
languages, in order to have access to an initiality theorem.
However, while doing so, we would like to preserve intuitions and results from signatures.
To do so, we see signatures with strength as particular signatures.

\begin{related Work}
  Signatures with strength \cite{FPT99,Hss04} have been introduced before
  modules over monoids and signatures \cite{HirschowitzMaggesi07,HirschowitzMaggesi12},
  and some authors solely rely on signatures with strength.
  However, modules over monoids are important conceptually, and some notions like
  presentable signatures \cite{PresentableSignatures21} or transition monads
  \cite{TransitionMonads22} do not seem to have clear counterparts in terms of
  strengths yet.
  It is hence paramount to relate both notions to unify the different approaches.
\end{related Work}

As identified in \cite{HirschowitzMaggesi12} for the particular monoidal
category $[\Set,\Set]$, every signature with strength yields a signature.
Indeed, signatures with strength can be seen as signatures where the module
substitutions have been particularly restricted:

\begin{proposition}
  \label{prop:sigstrength_to_sig}
  There is a functor $ι : \SigStrength(\mc{C}) → \Sig(\mc{C})$ from signatures with strength to signatures.
  It associates to any signature with strength $(H,θ)$ the signature (i.e., a functor)
  that associates to a monoid $(R,η,μ)$ the $R$-module $H(R)$ with
  the module substitution
  \[
    \begin{tikzcd}
      H(R) ⊗ R \ar[r, "θ_{R,η}"]
        & H(R ⊗ R) \ar[r, "H(μ)"]
        & H(R).
    \end{tikzcd}
  \]
  To a monoid morphism $f : R → R'$, it associates the morphism of modules $H(f) : H(R) → H(R')$.
  Given a morphism of signatures with strength $h ; (H,θ) → (H',θ')$,
  the functor associates the morphism of signatures $(\Id,h) : ι(H,θ) → ι(H',θ')$.
\end{proposition}

\begin{related Work}
  \label{related-work:model-sigma-monoids}
  Unfolding the definition of model (\cref{def:models}) for a signature with
  strength gives exactly a $Σ$-monoid, the notion of model used in
  \cite{FPT99,SecondOrderDep08,ListObjects17}.
  Indeed, a model is a monoid $(R,η,μ)$ with a morphism of module $r : Σ(R) → R$,
  in this case a morphism $r : Σ(R) → R$ in $\mc{C}$ such that the diagram
  \[
    \begin{tikzcd}
      H(R) ⊗ R \ar[r, "θ"] \ar[d, swap, "r ⊗ R"]
        & H(R ⊗ R) \ar[r, "H(μ)"]
        & H(R) \ar[d, "r"]\\
      R ⊗ R \ar[rr, swap, "μ"] & & R.
    \end{tikzcd}
  \]
  commutes. This is exactly the definition of a $Σ$-monoid.
  Consequently, for signatures with strength both approaches yield the same notion of
  model, and we can understand signatures with strength through the intuition of
  modules over monoids.
\end{related Work}

To understand signatures with strength through signatures
and use them instead to specify our languages, we show that the
functor $ι$ maps our basic constructions on signatures with strength, such as the trivial signature or
coproducts, to their counterpart in signatures.

\begin{proposition}
  The functor $ι : \SigStrength(\mc{C}) → \Sig(\mc{C})$ maps the trivial
  signature with strength $Θ : \SigStrength(\mc{C})$ and the $D ⊗ \_$
  construction to their counterpart.
\end{proposition}

\begin{proposition}[Preservation of colimits]
  \label{prop:sigstrength_sig_colimits-preserved}
  The functor $ι : \SigStrength(\mc{C}) → \Sig(\mc{C})$ preserves colimits
  provided that for all $B : \mc{C}$, $\_ ⊗ B$ preserves them.
\end{proposition}
\begin{proof}
  The image of the colimit is $(\Id + \colim H_i, μ ∘ \colim θ_i)$, and the
  colimit of the images is $(\colim (\Id + H_i), \colim(μ ∘ θ_i))$.
  Both are equal as colimits distribute over each other, that precomposition preserves colimits,
  and by the universal property of colimits.
\end{proof}

\begin{proposition}[Preservation of limits]
  \label{prop:sigstrength_sig_limits-preserved}
  The functor $ι : \SigStrength(\mc{C}) → \Sig(\mc{C})$ preserves limits
  provided that for all $B : \mc{C}$, $\_ ⊗ B$ preserves them, and that they
  distribute over binary coproducts.
\end{proposition}
\begin{proof}
  The proof is similar to that for colimits in \cref{prop:sigstrength_sig_colimits-preserved}.
  The extra assumption is needed as colimits do not always commute with limits
  \cite[Chapter 3.8]{CategoryTheoryInContext14}.
\end{proof}

We do not have a general result that $i$ preserves variable binding since the notion of variable binding depends on the monoidal category by which our framework is parametrized.
Consequently, when applying the framework one must also check that both representations of variable binding coincide.
Alternatively, as variable binding is a basic building block, one can also
simply directly define the module representation as the image of $i$.
Once checked, using the closure properties, we can prove that algebraic
signatures yield the same models.

\begin{example}
  The functor $ι : \SigStrength(\mc{[\Set,\Set]}) → \Sig(\mc{[\Set,\Set]})$ maps
  unary binding of signatures with strength $Θ^{(n)} : \SigStrength([\Set,\Set])$ to
  its signature counterpart $Θ^{(n)} : \Sig([\Set,\Set])$.
\end{example}

\begin{example}
  The functor $ι : \SigStrength(\mc{[\Set,\Set]}) → \Sig(\mc{[\Set,\Set]})$
  preserves algebraic signatures on $[\Set,\Set]$, like $Θ × Θ + Θ^{(1)}$
  specifying the untyped lambda calculus.
\end{example}

\subsection{The Initiality Theorem}
\label{subsec:initiality_theorem}

As for signatures, not all signatures with strength admit an initial model.
For instance, the signature $\mc{P} \circ Θ$ can be equipped with a
strength by postcomposition $Θ$ by $\mc{P} : \Set → ∖Set$ (\cref{ex:left-comp}).
Yet, it is not representable on $[\Set,\Set]$ by \cref{ex:not-representable}.
Thankfully, as signatures with strength $(H,θ)$ decompose into an
endofunctor $H : \mc{C} → \mc{C}$ and a strength $θ$, it is easier to
provide an initiality theorem and an adjoint theorem by requiring conditions
on $\mc{C}$ and $H$.
We now state both of these theorems, and prove them in \cref{sec:building_initial_model}

\begin{related Work}
  The following theorems are direct counterpart to results in
  \cite{FPT99,SecondOrderDep08,ListObjects17} up to the following technical details.
  First, compared to \cite{FPT99,SecondOrderDep08}, but akin to \cite{ListObjects17}
  we use $ω$-cocontinuity rather than monoidal closedness.
  Second, we deduce the adjoint theorem from the initiality theorem rather than
  the opposite, saving us the hypothesis that $X ⊗ \_$ is $ω$-cocontinuous in
  the initiality theorem.
  These changes are important for unifying the approaches and are discussed in
  detail in \cref{subsubsec:rw-co-vs-adj}.
\end{related Work}

\begin{restatable}[The Initiality Theorem]{theorem}{initialitytheorem}
  \label{thm:initiality-theorem}
  Let $\mc{C}$ be a monoidal category, with initial object, binary coproducts,
  $ω$-colimits and such that for all $Z : \mc{C}$, $\_ ⊗ Z$
  preserves initiality, binary coproducts and $ω$-colimits.
  Then, given a signature with strength $(H,θ)$, if $H$ is
  $ω$-cocontinuous, then the associated signature has an initial model
  $\ol{H}$, with the initial algebra $μ A.(I + H(A))$ as underlying object.
\end{restatable}

\noindent This theorem is very powerful as it provides a single initiality
theorem applicable to different input monoidal categories, hence handling
different kinds of contexts and different type systems.
Moreover, the conditions are relatively mild: in practice, we will always
work with cocomplete categories, and with nice monoidal products on the left.

Assuming additionally that for all $X : \mc{C}$, $X ⊗ \_$ preserves $ω$-colimits,
the initiality theorem can be extended to an adjoint theorem:

\begin{restatable}[The Adjoint Theorem]{theorem}{adjointtheorem}
  \label{thm:adjoint-theorem}
  Let $\mc{C}$ be a monoidal category, with initial object, binary coproducts,
  $ω$-colimits and such that for all $Z : \mc{C}$, $\_ ⊗ Z$
  preserves initiality, binary coproducts and $ω$-colimits.
  If additionally, for all $X : \mc{C}$, $X ⊗ \_$ preserves $ω$-colimits,
  then, for any $ω$-cocontinuous signature with strength $(H,θ)$,
  the forgetful functor $U : \Model(H) → \mc{C}$ has a left adjoint
  $\mrm{Free} : \mc{C} → \Model(H)$, where $\Free(X)$, for $X : \mc{C}$,
  has the initial algebra $μ A.(I + H(A))$ as underlying object.
\end{restatable}

In practice, one works with a fixed monoidal category $\mc{C}$ satisfying the
hypotheses of the initiality theorem, like $[\Set,\Set]$.
In that case, it suffices for a signatures with strength to be $ω$-cocontinuous
to be representable.

\begin{definition}
  $ω$-cocontinuous signatures with strength form a full subcategory $\SigStrength_ω(\mc{C})$
  of the category of signatures with strength.
\end{definition}

\begin{corollary}
  \label{coro:all-pres}
  Under the hypotheses of the initiality theorem, all signatures of
  $\SigStrength_ω(\mc{C})$ are representable, and there is a functor
  associating to each signature its initial model:
  \[ \ol{(\_)} : \SigStrength_ω(\mc{C}) \longrightarrow \mrm{SigModel}{(\mc{C})} \]
\end{corollary}

\noindent It is then particularly interesting to study their closure properties
to be able to specify languages and ensures they are representable modularly.
Thanks to the closure properties of signatures (\cref{prop:sigstrength-closure-colimits}),
it unfolds to the closure properties of $ω$-cocontinuity (\cref{subsec:omega-colimits})
that differ between limits and colimits.

\begin{proposition}[Closure under colimits]
  \label{prop:sigstrength_omega_cocomplete}
  By \cref{prop:omega-colimits,prop:sigstrength_sig_colimits-preserved}, if
  $\mc{C}$ is cocomplete and for all $B : \mc{C}$, $\_ ⊗ B$ preserves colimits,
  then the category $\SigStrength_ω(\mc{C})$ is cocomplete.
\end{proposition}

\begin{proposition}[Closure under limits]
  \label{prop:sigstrength_omega_complete}
  By \cref{prop:omega-limits,prop:sigstrength_sig_limits-preserved}, if $\mc{C}$
  admits a class of limits that commute with binary coproduct and $ω$-colimits,
  and that is preserved by $\_ ⊗ B$ for all $B : \mc{C}$, then the category
  $\SigStrength_ω(\mc{C})$ is closed under this class of limits.
\end{proposition}

For instance, consider the category $[\Set,\Set]$.
As it satisfies the hypotheses of the initiality theorem, any $ω$-cocontinuous signature with strength on $[\Set,\Set]$ is repesentable.
It hence suffices to specify languages as $ω$-cocontinuous signature with strength to have an initial model for them.
This is possible easily and modularly as they are are closed under sums and finite products.

\begin{proposition}
  Signatures with strength on $[\Set,\Set]$ $\SigStrength_ω([\Set,\Set])$
  have a terminal object, are closed under \emph{finite} products, and are
  closed under coproducts by
  \cref{prop:sigstrength_omega_cocomplete,prop:sigstrength_omega_complete,prop:presheaves-limits}.
\end{proposition}

\noindent It can easily be proven that the signature with strength $Θ^{(1)}$ is $ω$-cocontinuous.
From there, using the closure properties by coproducts, finite products and composition,
we can deduce that all algebraic signatures and usual untyped languages are representable.

\begin{example}
  All algebraic signatures (\cref{def:alg-sig,ex:alg-sig}) on $[\Set,\Set]$ are representable.
  In particular, the untyped lambda calculus specified by $Θ × Θ + Θ^{(1)}$, first
  order logic specified by $2Θ^0 + Θ + 3Θ^2 + 2Θ^{(1)}$, and linear logic specified
  by $2Θ^0 + 2Θ +5Θ^2 + 2Θ^{(1)}$ are representable.
\end{example}

\section{Building an Initial Model}
\label{sec:building_initial_model}

We have seen two different notions of signatures:
One the one hand, the signatures defined in terms of modules over monoids provide a flexible framework.
On the other hand, signatures with strength provide an initiality theorem and an adjoint theorem.
In this section, we prove these two theorems.
In particular, we will construct a ``syntactic'' model for any signature with
strength satisfying the hypotheses of \cref{thm:initiality-theorem}.

To do so, we rely on heterogeneous substitution systems (hss); these were originally
designed to prove that both well founded and non-well founded syntax \cite{Hss04}
have a monadic substitution structure, and extended to provide a form of initial
semantics \cite{HssRevisited15} in the well founded case.
We generalize the notion of hss from endofunctor categories to monoidal categories.
To prove the initiality theorem, we further generalize a proof scattered throughout
\cite{Hss04,DeBruijnasNestedDatatype99,HssRevisited15,HssUntypedUniMath19},
to monoidal categories and ω-cocontinuity, and adapt it from hss to models.
Doing so, we provide more details than currently available in the literature.

\begin{related Work}
  We stress that hss and a variant of this proof have been recently and
  independently generalized to monoidal categories by Matthes to ease
  formalization, as mentioned in \cite[Section 4.4]{HssNonWellfounded24}.
  We provide a full discussion on hss and this proof in \cref{subsec:rw-hss}.
\end{related Work}

\begin{related Work}
  Another proof following \cite{ListObjects17} would be possible.
  We explain the differences with this proof in details in
  \cref{subsubsec:rw-param-vs-hss,subsubsec:rw-building-param-hss}.
\end{related Work}

To convey intuiton, in the following, we first review hss and explain why
they allow us to construct models in \cref{subsec:hss_models}.
We then explain how to build a heterogeneous substitution system from our
assumptions in \cref{subsec:building_hss}, and why the model constructed from it
is initial as a model in \cref{subsec:building_initial_model}.
Finally, we prove the adjoint theorem from the initiality theorem.

In this section, we suppose that $\mc{C}$ has coproducts, and that they are
preserved by $\_ ⊗ Z$ for all $Z : \mc{C}$, as it is needed to state
our definitions.
As before, these hypotheses are by no means a restriction as they are also
hypotheses of the initiality theorem.

For readability, we write the functor $A ↦ I + H(A)$ as $I + H\_$ or $I + H$.

\subsection{Heterogeneous Substitution Systems and Models}
\label{subsec:hss_models}

The recursion principle arising from initial algebras is not sufficient to
prove that a higher-order language, such as the lambda calculus, forms a monad.
As explained, for $[\Set,\Set]$, in \cite{DeBruijnasNestedDatatype99,GeneralisedFold99},
a stronger recursion principle is required; in their case, one named ``generalized fold''.
Heterogeneous substitution systems (hss), introduced in \cite{Hss04,HssRevisited15}
on endofunctor categories, are inspired by generalized folds, and are
designed to prove that higher-order languages form monads when considered
with their substitutions.

\begin{definition}[Heterogeneous Substitution System]
  \label{def:hss}
  For a signature with strength $(H, θ)$, a \emph{heterogeneous substitution
  system (hss)} is a tuple $(R, η, r)$ where $R : \mc{C}$ is an object
  of $\mc{C}$ and $η : I → R$ and $r : H(R) → R$ are morphisms
  of $\mc{C}$ such that, for all $(Z , e) : \trm{Ptd}(\mc{C})$ and $f : Z
  → R$, there is a unique morphism $\{ f \} : R ⊗ Z →
  R$ making the following diagram commute:
  \[
    \begin{tikzcd}[column sep=large]
      I ⊗ Z \ar[r, "η ⊗ Z"] \ar[dd, swap, "λ_Z"]
        & R ⊗ Z \ar[dd, "\{f\}"]
        & H(R) ⊗ Z \ar[l, swap, "r ⊗ Z"]
                         \ar[d, "θ_{R,e}"] \\
        &
        & H(R ⊗ Z) \ar[d, "H(\,\{f\}\,)"] \\
      Z \ar[r , swap, "f"]
        & R
        & H(R) \ar[l, "r"]
    \end{tikzcd}
  \]
\end{definition}

\begin{related Work}
  In previous works on hss \cite{Hss04,HssRevisited15,HssUntypedUniMath19,HssTypedUnimath22},
  it was required for $f$ to be a morphism of pointed objects $f : (Z,e) → (R,η)$.
  It was later realised independently by Matthes, Wullaert, and Ahrens in \cite{HssNonWellfounded24},
  and by Lafont that this hypothesis is superfluous for our purpose.
\end{related Work}

Intuitively, hss strengthen the concept of algebra as it additionally enables us
to build a unique morphism $R ⊗ Z → R$ for any suitable $Z$.
As done in \cite{Hss04} for endofunctor categories, the added freedom of choosing
$Z$ as we wish enables us to derive a monoid structure, and later a model structure:

\begin{proposition}[Monoids from Hss]
  \label{prop:hss_to_monoid}
  If a signature $(H, θ$) has a heterogeneous substitution system $(R,
  η, r)$, then $(R, η, \{ \id_{(R,η)} \})$ is a monoid.
\end{proposition}
\begin{proof}
  Let's denote $\{\id_{(R,η)}\} : R ⊗ R → R$ by $μ$.
  By definition, $(R, η, μ)$ is well-typed.
  Moreover, the left-unit law of the monoid $λ_R = μ ∘ (η ⊗ R)$
  holds by definition of $μ$.
  Hence, there are only the right-unit law and the associativity of $μ$
  to check.

  To prove the second unit law, $\rho_R = μ ∘ (R ⊗ η)$, let's
  apply the hss for $f := η : I → R$.
  Both morphisms make the diagram commute, thus are equal by uniqueness.
  The morphism $\rho_R$ can be shown to satisfy the diagram using that
  $λ_I = \rho_I$, the definition of $θ_{R,\id}$, and the
  naturality of $\rho$.
  The morphism $μ ∘ (R ⊗ η)$ can be shown to satisfy the
  diagram using the left-unit law, the definition of $μ$, and the
  naturality of $λ$, and of $θ$ in the second argument.
  Thus by uniqueness $\rho_R = \{ η \} = μ ∘ (R ⊗ η)$.

  For homogeneity reasons, to prove the associativity of $μ$, we prove $μ
  ∘ (R ⊗ μ) = μ ∘ (μ ⊗ R) ∘ \alpha^{-1}$.
  To do so, let's apply the hss in $μ : R ⊗ R → R$, where $R ⊗ R$ is pointed by
  $I \xrightarrow{λ_I} I ⊗ I \xrightarrow{η ⊗ η} R ⊗ R$.
  Both morphisms make the diagram commute, thus are equal by uniqueness.
  The morphism $μ ∘ (μ ⊗ R)$ can be shown to satisfy it using
  the unit law, the naturality of $θ$, and the definition of $μ$.
  The morphism $μ ∘ (μ ⊗ R) ∘ \alpha^{-1}$ can be shown to
  satisfy the diagram using the naturality of $θ$ and $\alpha$, the
  definition of $μ$, and the definition of $θ$ in $e ⊗ e'$.
\end{proof}

\begin{proposition}[Models from Hss]
  \label{prop:hss_to_model}
  If a signature $(H, θ$) has a heterogeneous substitution system
  $(R,η,r)$, then $((R,η,μ), r)$ is a model of $H$.
\end{proposition}
\begin{proof}
  By \cref{prop:hss_to_monoid}, $(R,η,μ)$ is a monoid.
  It remains to prove that $r$ is a morphism of modules $H(R) → R$ i.e.
  to prove that $μ ∘ (r ⊗ R) = r ∘ H(μ) ∘ θ_{R,\id}$.
  This is true by the definition of $μ$.
\end{proof}

\subsection{Building Heterogeneous Substitution Systems}
\label{subsec:building_hss}

\Cref{prop:hss_to_model} states that it suffices to build a heterogeneous
substitution system for a signature with strength to get a model for it.
In this section, we build a heterogeneous substitution system from a signature
with strength $(H, \Theta)$, using generalized iteration in Mendler's style.
As the construction relies on a particular way of building initial algebras, we
will first recall, and give a proof of, Adámek's Theorem.

\begin{theorem}[Adámek's Theorem \cite{Adamek74}]
  \label{thm:adamek}
  Let $\mc{C}$ be a category with an initial object $0$, and
  $\omega$-colimits. Let $F : \mc{C} → \mc{C}$ be an $\omega$-cocontinuous
  functor.
  Then there is an initial $F$-algebra $(R,r)$ with $R$ built as the
  $\omega$-colimit of:
  \[
    \begin{tikzcd}[column sep=large]
      0 \ar[r, "i"] \ar[dr, swap, "t_0"]
        & F(0) \ar[r, "F(i)"] \ar[d, "t_1"]
        & F^2(0) \ar[r, "F^2(i)"] \ar[dl, "t_2"]
        & ... \\
      & R
        &
        &
        &
    \end{tikzcd}
  \]

\end{theorem}
\begin{proof}
  As $F$ preserves $\omega$-colimits, $F(R)$ is the colimit of the image by
  $F$ of the chain $F(\trm{chn}_F)$.
  Using the initiality of $0$, it can be completed into a cocone for the
  full chain $\trm{chn}_F$.
   Given any other cocone $A$ of $\trm{chn}_F$, $A$ is, in particular, a
   cocone for $F(\trm{chn}_F)$, which yields a map $h : F(R) → A$
   satisfying the universal property of $F(\trm{chn}_F)$.
   Using initiality, $h$ also verifies the one of $F(\trm{chn}_F)$.
  Given any other such map $h'$, in particular both $h$ and $h'$ satisfy the
  universal property $F(\trm{chn}_F)$, hence as $F(R)$ is a colimit of this
  chain, $h = h'$.
  Thus $F(R)$ is a colimit of $\trm{chn}_F$, which yields a unique map
  $r : F(R) → R$.

  Given an $F$-algebra $(A,a)$, we build a cocone on $A$.
  We define $a_n$ recursively: the morphism $a_0$ is the unique morphism
  from $0 → A$; for a natural number $n$, $a_{n+1}$ is defined as the
  composition $F^{n+1}(0) \xrightarrow{a_n} F(A) \xrightarrow{a} A$.
  The commutativity holds by the uniqueness in the $0$ case and recursively
  for $n+1$.
  Then by the universal property we obtain a unique map $h : C → A$.
  Both $a∘ F(h)$ and $h ∘ c$ satisfy the universal property of the
  colimit, as such they are equal and so $h$ is a morphism of algebras.
  Given another morphism of $F$-algebras $h'$, it can be shown by induction
  that it verifies the universal property, hence by uniqueness $h = h'$, and
  so $h$ is unique.
  In consequence, $(R,r)$ is an initial $F$-algebra.
\end{proof}

\begin{restatable}[Generalized Mendler's style Iteration {{\cite[Theorem 1]{GeneralisedFold99}}}]{theorem}{Mendler}
  \label{thm:gen-mendler}
  Let $\mc{C}$ and $\mc{D}$ be two categories with initial object and
  $\omega$-colimits, $F : \mc{C} → \mc{C}$ and $L : \mc{C} → \mc{D}$ be
  two functors.
  If $F$ and $L$ preserve $\omega$-colimits, and $L$ preserves initiality,
  then for all $X : \mc{D}$ and natural transformation
  \[ Ψ : \mc{D}(L\_,X) → \mc{D}(L(F\_),X) \]
  there is a unique morphism $\mrm{It}^L_F(Ψ) : L(R) → X$ in $\mc{D}$
  --- where $(R,r)$ is the initial $F$-algebra obtained by Adámek's Theorem (\cref{thm:adamek}) ---
  making the following diagram commute:
  \[
    \begin{tikzcd}[ampersand replacement=\&]
      L(F(R)) \ar[r, "L(r)"] \ar[dr, swap, "Ψ_R(\mrm{It}^L_F(Ψ))"]
        \& L(R) \ar[d, "\mrm{It}^L_F(Ψ)"] \\
      \& X
    \end{tikzcd}
  \]
\end{restatable}
\begin{proof}
  As a shorthand, we will denote $\mrm{It}^L_F(Ψ)$ by $h$ in the proof of existence and uniqueness of $\mrm{It}^L_F(Ψ)$.
  First, let's construct $h$.
  As $L$ preserves initiality, $L(0)$ is initial and there is a unique map
  $x : L(0) → X$.
  Iterating $Ψ$ yields a cocone $(X, (Ψ^n(x))_{n : \mb{N}})$ for the
  diagram $(LF^n(0), LF^n(i))_{n : \mb{N}}$.
  Yet as $L$ preserves $\omega$-colimits, $L(R)$ is the colimit of this
  diagram, and by the universal property of colimits, there is a unique $h :
  R(L) → X$ such that $\forall n.\; h ∘ L(t_n) = Ψ^n(x)$, i.e.,
  such that the following diagram commutes:
  \[
    \begin{tikzcd}[column sep=large]
      L(0) \ar[r, "L(i)"] \ar[dr, swap, "L(t_0)"] \ar[ddr, bend right, swap, "Ψ^0(x)"]
        & LF(0) \ar[r, "LF(i)"] \ar[d, "L(t_1)"]
        & LF^2(0) \ar[r, "LF^2(i)"] \ar[dl, "L(t_2)"] \ar[ddl, bend left, "Ψ^2(x)"]
        & ... \\
      & L(R) \ar[d, "h"]
        &
        &
        & \\
      & X
        &
        &
        &
    \end{tikzcd}
  \]
  We have built a map $h$, it remains to prove that it satisfies the
  desired property:  $h ∘ L(r) = Ψ(h)$.
  To do so we are going to prove that $\forall n.\; Ψ(h) ∘
  L(r^{-1}) ∘ L(t_n) = Ψ^n(x)$.
  Indeed by the uniqueness of $h$, we can conclude that $h = Ψ(h) ∘
  L(r^{-1})$, i.e. $h ∘ L(r) = Ψ(h)$.
  We prove $\forall n.\; Ψ(h) ∘ L(\alpha^{-1}) ∘ L(t_n) =
  Ψ^n(x)$ by induction on $n$.
  The $n = 0$ case holds by initiality of $L(0)$.
  The $n+1$ case holds by the following chain of equalities:
  \[
  \begin{array}{cclc}
    Ψ(h) ∘ L(r^{-1}) ∘ L(t_{n+1})
      &=& Ψ(h) ∘ L(r^{-1}) ∘ L(r) ∘ LF(t_{n})
          & (\textrm{definition of }r) \\
      &=& Ψ(h) ∘ LF(t_{n}) & \\
      &=& Ψ(h ∘ L(t_{n})) & (\textrm{naturality of }Ψ) \\
      &=& Ψ(Ψ^n(x)) & (\textrm{induction hypothesis})
  \end{array}\]
  Now that we have proven existence, we need to prove uniqueness.
  To do so, suppose we have a $h : L(R) → X$ such that $h ∘ L(r) =
  Ψ(h)$; we are going to show that $\forall n.\; h ∘ L(t_n) =
  Ψ^n(x)$.
  As a consequence, such an $h$ verifies the universal property of $L(R)$ and
  hence $h$ is unique. We prove the equations by induction on $n$.
  The $n = 0$ case holds by initiality of $L(0)$.
  The $n+1$ case holds by the following chain of equation:
  \[
  \begin{array}{cclc}
    h ∘ L(t_{n+1})
      &=& h ∘ L(r ∘ t_n)
          & (\textrm{definition of }r) \\
      &=& h ∘ L(\alpha) ∘ L(t_n) & \\
      &=& Ψ(h) ∘ L(t_n) & (\textrm{definition of }h) \\
      &=& Ψ(h ∘ LF(t_n)) & (\textrm{naturality of }Ψ) \\
      &=& Ψ(Ψ^n(x)) & (\textrm{induction hypothesis})
  \end{array}\]
\end{proof}

\begin{proposition}[Building an Hss]
  Let $\mc{C}$ be a monoidal category, with initial object, coproducts,
  $\omega$-colimits and such that for all $Z : \mc{C}$, $\_ ⊗ Z$
  preserves initial objects, coproducts, and $\omega$-colimits.
  Let $(H,θ)$ be a signature with strength such that $H$ is
  $\omega$-cocontinuous.
  Then $(R, η + r)$ --- the initial algebra of $I + H\_$ obtained by
  Adámek's Theorem --- is a heterogeneous substitution system for
  $(H,θ)$.
\end{proposition}

\begin{proof}
  Let $(Z,e)$ be a pointed object, and $f : Z → R$ a morphism.
  We are going to construct $\{ f \}$ and show that it is uniquely defined,
  by a suitable application of generalized Mendler's style iteration.
  Let's apply the theorem for $F := \id + H$, $L := \_ ⊗ Z$, $X := R$
  and $Ψ \;h ↦ (f + r) ∘ (\id + H(h)) ∘ (λ_Z +
  θ_{R,e})$.
  This will give us a unique map $\mrm{It}^{\_ ⊗ Z}_{I + H}(Ψ) : R
  ⊗ Z → R$, such the following diagram commutes:
  \[
  \begin{tikzcd}[column sep=4cm]
    I ⊗ Z + H(R) ⊗ Z
                        \ar[r, "(η + r) ⊗ Z"]
                        \ar[d, swap, "λ_Z + θ_{R,e}"]
                        \ar[ddr, start anchor=-25, shorten <=8pt, end anchor=north west,
                             "Ψ(\mrm{It}^{\_ ⊗ Z}_{I + H}(Ψ))"]
      & R ⊗ Z     \ar[dd, "\mrm{It}^{\_ ⊗ Z}_{I + H}(Ψ)"]\\
    Z + H(R ⊗ Z)  \ar[d, swap, "\id + H(\mrm{It}^{\_ ⊗ Z}_{I + H}(Ψ))"]
      & \\
    Z + H(R)            \ar[r, swap, "{[}f {,} r{]}"]
      & R \\
  \end{tikzcd}
  \]
  This diagram is exactly the definition of an hss, if one denotes
  $\mrm{It}^{\_ ⊗ Z}_{I + H}(Ψ)$ by $\{ f \}$.
  Hence it suffices to verify the hypothesis for the given input to get the
  result.
  The only non-obvious hypothesis is that the above defined $Ψ$ is natural
  in $R$, which follows by naturality of $θ$ in its first argument.
\end{proof}

\subsection{Building an Initial Model}
\label{subsec:building_initial_model}

The previous work has enabled us to construct a model of a given signature.
It remains to prove that this model is initial.
To do so, we adapt a proof for hss on endofunctor categories introduced in
\cite{HssRevisited15} to models on monoidal categories.
This crucially relies on a fusion law for generalized Mendler's style iteration
which enables us to factorise a generalized Mendler's style iteration followed
by a natural transformation into a Mendler's style iteration.
This is not surprising as we are relying on a generalized recursion principle
to build our model, and fusion laws are ubiquitous in computer science to
simplify recursion principles.

\begin{theorem}[Fusion Law for Generalized Mendler's style Iteration {{\cite[Lemma 9]{HssRevisited15}}}]
  \label{thm:fusion-law}
  Let $\mc{C,D},F,L, X, Ψ$ be objects satisfying the hypotheses of the
  \hyperref[thm:gen-mendler]{generalized Mendler's style iteration} theorem.
  For the same $\mc{C,D},F$, suppose given other $L',X',Ψ'$ satisfying
  the hypotheses.
  If there is a natural transformation $Φ : \mc{D}(L\_,X) →
  \mc{D}(L'\_,X')$ such that $Φ_{F(μ F)} ∘ Ψ_{μ F} =
  Ψ'_{μ F} ∘ Φ_{μ F}$ --- where $μ F$ denotes the
  $F$-algebra built by Adámek's theorem --- then
  \[ Φ_{μ F}(\mrm{It}^L_F(Ψ)) = \trm{It}^{L'}_F(Ψ') \]
\end{theorem}
\begin{proof}
  By uniqueness, it suffices to prove that $Φ_{μ F}(\trm{It}^L_F(Ψ))$
  satisfies the defining diagram of $\trm{It}^{L'}_F(Ψ')$.
  This can be done using the definition of the assumption, the definition of
  $\trm{It}^L_F(Ψ)$ and the naturality of $Φ$.
\end{proof}

We are now ready to prove the initiality theorem:

\initialitytheorem*
\begin{proof}
  By \cref{prop:hss_to_model}, $((R,η, μ), r)$ is a model of $H$.
  We need to prove that this model is actually initial.
  Let $((R',η', μ'), r')$ be another model.
  We need to prove there is a unique morphism of models $((R,η, μ),
  r) → ((R',η',μ'),r')$.
  In other words, we need to prove there is a unique morphism of monoids $f
  : (R,η,μ) → (R',η',μ')$ --- i.e., respecting $η$ and $μ$
  --- making the following diagram commute:
  \[
    \begin{tikzcd}
      H(R) \ar[r, "r"] \ar[d, swap, dashed, "H(f)"]
        & R \ar[d, dashed, "\exists ! f"] \\
      f^*H(R') \ar[r, swap, "f^*r'"] & f^*R'
    \end{tikzcd}
  \]
  To prove both uniqueness and existence, we are going to use that $(R',
  η' + r')$ is a $(I + H\_)$-algebra and that $(R, η + r)$ is the
  initial one. \medskip

  To prove uniqueness, suppose we have a morphism of models $f$, in
  particular it is a morphism of $(I + H\_)$-algebras $f : (R, η + r) →
  (R', η' + r')$ and as such is unique by the initiality of $(R, η
  + r)$. \medskip

  To prove the existence of such a model morphism, we are going to show that
  the morphism of algebras $f : (R, η + r) → (R', η' + r')$
  existing by initiality of $(R, η + r)$ is a morphism of models.
  The morphism $f$ respects $η$ as it is a morphism of algebras.
  The commutativity of the diagram of module morphisms holds if it holds for
  the underlying morphism of $\mc{C}$, which is also verified as $f$ is a
  morphism of algebras.

  It remains to prove that it respects $μ$, i.e., that $f ∘ μ = μ'
  ∘ f ⊗ f$.
  To do so, we are going to use that $μ := \{ \id \} := \mrm{It}^{\_
  ⊗ Z}_{I + H}(Ψ)$  to factorise $f ∘ μ$ into another
  iteration $\mrm{It}^{\_ ⊗ Z}_{I + H}(Ψ')$ as shown below.
  The point is that iterations are the unique morphisms making their
  associated diagrams commute.
  In consequence, it will suffice to prove that $μ' ∘ f ⊗ f$
  satisfies it to prove the equalities.
  \begin{align*}
    \begin{tikzcd}[ampersand replacement=\&]
      R ⊗ R \ar[r, "f ⊗ f"]
                  \ar[d, swap, "\mrm{It}^{\_ ⊗ Z}_{I + H}(Ψ)"]
        \& R' ⊗ R' \ar[d, "μ'"] \\
      R \ar[r, swap, "f"]
        \& R'
    \end{tikzcd}
    &&
    \begin{tikzcd}[ampersand replacement=\&]
      R ⊗ R \ar[r, "f ⊗ f"] \ar[dr, swap, bend right, "\mrm{It}^{\_ ⊗ Z}_{I + H}(Ψ')"]
        \& R' ⊗ R' \ar[d, "μ'"] \\
        \& R'
    \end{tikzcd}
  \end{align*}
  Here $Ψ \;h := (\id + r) ∘ (\id + H(h)) ∘ (λ_Z +
  θ_{R,e})$ and $Ψ' \;h := (f + r') ∘ (\id + H(h)) ∘
  (λ_Z + θ_{R,e})$.
  To do the factorisation, we apply the fusion law for $Φ := f^*$ i.e.
  precomposition by $f$, for which the assumption is satisfied as $f$ is a
  morphism of algebras.
  Finally, proving that $μ' ∘ f ⊗ f$ satisfies the diagram of
  $\mrm{It}^{\_ ⊗ Z}_{I + H}(Ψ')$ unfolds to proving:

  \begin{align*}
    \begin{tikzcd}[ampersand replacement=\&]
      I ⊗ R \ar[ddd, swap, "λ_R"] \ar[rr, "η ⊗ R"]
                  \ar[dr, swap, "I ⊗ f"]
        \&
        \& R ⊗ R \ar[d, "R ⊗ f"] \\
        \& I ⊗ R' \ar[r, "η ⊗ R'"]
                       \ar[dr, swap, near end, "η' ⊗ R'"]
                       \ar[ddr, swap, bend right, near start, "λ_{R'}"]
        \& R ⊗ R' \ar[d, "f ⊗ R'"] \\
        \&
        \& R' ⊗ R' \ar[d, "μ'"] \\
      R \ar[rr, "f"]
        \&
        \& R'
    \end{tikzcd}
    &&
    \begin{tikzcd}[ampersand replacement=\&]
      H(R) ⊗ R \ar[rr, "r ⊗ R"] \ar[dr, "H(f) ⊗ f"]
                     \ar[d, swap, "θ_{R,\id}"]
        \&
        \& R ⊗ R \ar[dd, "f ⊗ f"] \\
      H(R ⊗ R) \ar[d, swap, "H(f ⊗ f)"]
        \& H(R') ⊗ R' \ar[dl, "θ_{R',\id}"] \ar[dr, "r' ⊗ R'"]
        \& \\
      H(R' ⊗ R') \ar[d, swap, "H(μ')"]
        \&
        \& R' ⊗ R' \ar[d, "μ'"]\\
      H(R') \ar[rr, swap, "r'"]
        \&
        \& R'
    \end{tikzcd}
  \end{align*}
  The left diagram commutes using the naturality of $λ$, the fact that
  $f$ is a morphism of algebras, and by the monoid laws.
  The right diagram commutes by naturality of $θ$ in both arguments,
  the fact that $f$ is a morphism of algebras and because $r$ is a
  morphism of modules.
\end{proof}

\subsection{Building an Adjoint}
\label{subsec:building-adjoint}

While it is possible to directly prove the adjoint theorem (\cref{thm:adjoint-theorem}), and deduce the
initiality theorem (\cref{thm:initiality-theorem}) from it, it is possible and actually better to do the
opposite.
Indeed, by doing so, one can save, in the initiality theorem, the hypothesis
that $X ⊗ \_$ is $\omega$-cocontinuous:

\adjointtheorem*
\begin{proof}
  First, let's prove the existence of $\Free : \mc{C} → \Model(H)$.
  Given $X : \mc{C}$, $X ⊗ \_$ can be equipped with the strength
  $\alpha : (X ⊗ A) ⊗ B → X ⊗ (A ⊗ B)$.
  Hence, by \cref{prop:sigstrength_omega_cocomplete}, $H + X ⊗ \_$
  is an $\omega$-cocontinuous signature with strength, and so has an initial
  model by the initiality theorem denoted $\Free(X)$, with carrier $μ A.
  (I + H(A) + X ⊗ A)$.
  We then get a model of $H$ by forgetting the extra $X ⊗ \_$ structure.
  Given a morphism $X → Y$, $\Free(Y)$ can be equipped with a $H + X
  ⊗ \_$ structure.
  Hence, by initiality, there is a unique $H + X ⊗ \_$ morphism of
  model $\Free(X) → \Free(Y)$, which is in particular a $H$ morphism of
  models.

  Given an object $X : \mc{C}$, and a model of $H$, $M : \Model(H)$, we need
  to build an natural isomorphism $\Model(H)(\Free(X),\, M) \cong C(X,\,
  M)$, where we identify $U(M)$ with $M$.
  To build $K : C(X,\, M) → \Model(H)(\Free(X),\, M)$, given $f : X →
  M$, we turn $M$ from an $H$ model to an $H + X⊗ \_$ model.
  Indeed, by initiality of $\Free(X)$, this will provide a unique morphism
  of $H + X ⊗ \_$ models, which is in particular a morphism of
  $H$-models.
  To do so, it suffices to provide a module morphism $X ⊗ \Theta →
  \Theta$, which can be defined as $X ⊗ M \xrightarrow{f ⊗ M} M
  ⊗ M \xrightarrow{\mu_M} M$.
  We can build an inverse $L : \Model(H)(\Free(X),\, M) → C(X,\, M)$ using
  that $\Free(X)$ is a $H + X ⊗ \_$ model, and as such is equipped
  with morphism $\sigma : X ⊗ \Free(X) → \Free(X)$.
  Given $f^\# : \Free(X) → M$, we define $L(f^\#)$ as
  $X \xrightarrow{\rho_X}         X ⊗ I
     \xrightarrow{X ⊗ η} X ⊗ \Free(X)
     \xrightarrow{\sigma}         \Free(X)
     \xrightarrow{f^\#}           M$.
  Proving that $L ∘ K (f) = f$, follows the universal property of $G(f)$,
  the definition of $L$ and the monoid laws.
  By definition, $K(L(f^\#))$ is the unique morphism of $H + X ⊗ \_$
  models from $X$ to $M$ when equipped with $L(f)$.
  Hence, to prove that $K(L(f^\#)) = f^\#$, it suffices to prove that $f^\#$
  is such a morphism.
  As $h$ is already a morphism of $H$ models, it suffices to verify that it
  is compatible with the $X ⊗ \_$ constructor.
  It follows by the definition of the strength $X ⊗ \_$ and the
  different monoid and monoidal laws.
  Lastly, the naturality in both arguments follows from the definition.
\end{proof}

\section{Related Work}
\label{sec:related-work}

The literature on initial semantics is quite prolific, and contains many
different approaches to initial semantics.
We focus on three traditions, which we name after the different mathematical structures used in each of them: $Σ$-monoids, modules over monads, and heterogeneous
substitution systems.
The links between these different approaches are not, in our view, clearly laid out in the literature.
Furthermore, some of the traditions contain several different variations, which can be
confusing when trying to learn or use results from the literature.
To bridge this gap in the literature, we have suitably abstracted and combined
the different approaches in
\cref{sec:models,sec:initiality_theorem,sec:building_initial_model}.
This enables us to shed light on the prolific literature: by relating the
different existing approaches to our framework, we also relate them to each other.

Note that in this work, we do not tackle initial semantics for metavariables,
equations, reductions, or formalization in computer proof assistants.
Consequently, in this section on related work, we mention work on these subjects
but do not discuss them in detail.

We first give a brief, and necessarily incomplete, chronological overview of the
different lines of work on initial semantics in \cref{subsec:rw-overview}.
We then provide an extensive and detailed discussion tradition by tradition,
where we survey the different variations, the links with our framework, and
justify the design choices of our framework.
We discuss work on $Σ$-monoids in \cref{subsec:rw-sigma-mon}, on modules over
monoids in \cref{subsec:rw-modules-over-monoids}, and on heterogeneous
substitution systems in \cref{subsec:rw-hss}.

\subsection{Overview of Initial Algebra Semantics for Languages With Variable Binding}
\label{subsec:rw-overview}

In this section, we provide a chronological overview of the literature on initial semantics.
To enhance readability, the overview is divided in three parts.
These divisions reflect our understanding of the evolution of the literature,
and the involvement of new contributors, e.g. new PhD students.
Note that any such segmentation is inherently arbitrary and does not capture all the nuances.

\subsubsection{Origins: Up until 2007}

The concept of using monads as an abstraction to reason about simultaneous
substitution was first introduced by Bellegarde and Hook \cite{BellegardeHook94}.
It was generalized to using monoids in a monoidal category by Fiore, Plotkin, and Turi \cite{FPT99}.
To ease the construction of the substitution monad, building on the concept of
``nested'' data types \cite{NestedDataTypes98}, Bird and Paterson
\cite{DeBruijnasNestedDatatype99} defined the untyped lambda calculus
intrinsically, and strengthened the usual fold operation to ``generalized
folds''.
The authors then generalized the concept of ``generalized fold'' to generic
nested data types in \cite{GeneralisedFold99}.
As an alternative to generalized folds, Altenkirch and Reus considered instead
structural induction for nested data types in \cite{AltenkirchReus99}.

Capturing the substitution structure and recursion of an untyped higher-order
language as an initial object was first achieved in \cite{FPT99} by Fiore,
Plotkin, and Turi, using the category $[\mb{F},\Set]$.
To do so, they introduced $Σ$-monoids, and suggested using strength and monoidal
categories to prove initiality results.
Fiore then investigated simply-typed higher-order languages using those
methods in \cite{Cbn02,MMCCS05}.

Matthes and Uustalu introduced heterogeneous substitution systems on endofunctor
categories in \cite{Hss04}, in order to prove that both wellfounded and
non-wellfounded syntax have a monadic substitution structure.
In particular, they introduced signatures with strength as a formal notion
of signature for the first time, and proved generic theorems using them.

The framework based on monoidal categories and $Σ$-monoids was later detailed
by Fiore and Hamana, in \cite{SecondOrderDep08}, were they furthermore
considered the addition of meta-variables, and suggested a method to handle
dependently typed languages.
We do not deal with meta-variables in this work.

Modules over monads and associated models were introduced for endofunctor
categories by André Hirschowitz and Maggesi in \cite{HirschowitzMaggesi07}, in
order to capture the substitution properties of constructors.
This work was extended in \cite{HirschowitzMaggesi10}, were they proved, using a
proof specific to the base category of sets, that untyped higher-order languages
have an initial model on $[\Set,\Set]$.

\subsubsection{Consolidation: 2007 - 2015}

Equations for $Σ$-monoids were considered in Hur's dissertation \cite{HurPhd}
with Fiore, and reported in \cite{FioreHur07,FioreHur09,FioreHur08}.
Equations and meta-variables have also been considered in \cite{FioreMahmoud10,FioreHur10}.
We do not deal with semantics in this work, neither with denotational semantics in the form of equations, nor with operational semantics in the form of reduction rules.

The links between the $Σ$-monoids approach and the module over monads one
were investigated in Zsidó dissertation \cite{ZsidoPhd10}.
In particular, Zsidó fully worked out the simply-typed instance for both traditions.
A variant of Zsidó's construction for modules over monads was formalized with
Ahrens \cite{ISCoq10} in the proof assistant Coq.

A proposition to handle polymorphic languages was investigated by Hamana in
\cite{Polymorphism11}, and considered with meta-variables and equations
by Fiore and Hamana in \cite{PolymorphismEq13}.

The current concept of signatures for modules over monads was introduced by
Hirschowitz and Maggesi in \cite{HirschowitzMaggesi12}, and they showed that
every signature with strength induce a signature.

Modules over monads for syntax and semantics were investigated further in
Ahrens' dissertation \cite{AhrensPhd}.
He developed a framework to extend the initiality principle of
simply-typed languages to allow for translations across typing systems
\cite{ExtendedInitiality12}.
Ahrens also considered reductions rules using modules over monads and relative monads.
A framework for untyped languages was published in \cite{UntypedRelativeMonads16},
and for simply-typed languages in \cite{TypedRelativeMonads19}.

Heterogeneous Substitution Systems have been revisited in \cite{HssRevisited15},
by Ahrens and Matthes, where they extend the hss framework with an initiality
result and a formalisation.

$Σ$-monoids have also been revisited by Fiore and Saville in \cite{ListObjects17}.
They extended $Σ$-monoids to $T$-monoids, weakened the assumptions of
previous theorems, and provided more detailed proofs of the theorems.

\subsubsection{Recent Work: 2018 - Present}

Modules over monads and semantics in the form of equations and reductions between terms were investigated in
Ambroise Lafont's dissertation \cite{LafontPhd}, in collaboration
with other people, in particular, Ahrens, André Hirschowitz, Tom Hirschowitz, and Maggesi.
Equations were investigated through signatures in \cite{PresentableSignatures21},
and on their own in \cite{2Signatures19}.
Reductions rules and reduction strategies were investigated in
\cite{ReductionMonads20} and \cite{TransitionMonads20,TransitionMonads22}, respectively.

Heterogeneous substitution systems were applied by Ahrens, Matthes, and
Mörtberg to handle untyped and simply-typed higher-order languages in Coq's
UniMath library in \cite{HssUntypedUniMath19,HssTypedUnimath22}.

The framework using $Σ$-monoids was extended by Borthelle, Lafont, and
Tom Hirschowitz to skew-monoidal categories in \cite{CellularHoweTheorem20},
in order to study bisimilarity.

Another skew-monoidal category was investigated by André and Tom
Hirschowitz, Lafont and Magessi in \cite{NamelessDummies22}, to study
De Bruijn monads.

Fiore and Szamozvancev used the work on $Σ$-monoids,
meta-variables, and equation to design a framework to handle higher-order
languages in Agda \cite{FioreSzamozvancevPopl22}.

Very recently and independently from us, heterogeneous substitution systems and
results about them have been generalized to monoidal categories in the study of
non-wellfounded syntax in \cite{HssNonWellfounded24}.

\subsection{$Σ$-monoids}
\label{subsec:rw-sigma-mon}

\subsubsection{Origins}

Capturing higher-order languages with their substitution structure as initial
models was first achieved for untyped algebraic signatures \cref{def:alg-sig},
in a seminal work by Fiore, Plotkin and Turi in \cite{FPT99}, using the category
$[\mb{F},\Set]$.
They first showed that the pure syntax of untyped higher-order calculi, specified by
algebraic signatures (\cref{def:alg-sig}), is modeled by particular algebras on $[\mb{F},\Set]$.
They then introduced $Σ$-monoids, a particular case of model models for signatures
with strength by \cref{related-work:model-sigma-monoids}, to capture substitution.
They further stated that every binding signature induces an initial $Σ$-monoids
on $[\mb{F},\Set]$, thus providing a framework for initial semantics.
No proofs were given in the extended abstract \cite{FPT99}; but, the authors
suggested it can be proven using that, since $[\mb{F},\Set]$ is a
\emph{closed monoidal category}\footnote{
  A monoidal category is (bi)closed, when for all $Z : \mc{C}$, the functors
  $\_ ⊗ Z$ and $Z ⊗ \_$ have right adjoints.
} every free $Σ$-algebra such that $Σ$ has a pointed strength is parametrically free.

This claim was made more precise by Fiore in another extended abstract
\cite[Sections I.1.1 - I.2.2]{SecondOrderDep08}.
After giving an analysis of (pointed) strength in terms of actions over monoidal
categories, Fiore stated that if a monoidal category is closed and has binary
products, then every $(I + Σ + X ⊗ \_)$-initial algebra, such that $Σ$ has a
pointed strength, yields by parametrized initiality the free $Σ$-monoids over $X$.
Fiore furthermore suggested that if all of the $(I + Σ + X ⊗ \_)$-initial algebras
exist,\footnote{
  This is the case as soon as the monoidal category additionally has initial
  object and $ω$-colimits, and $Σ$ and $X ⊗ \_$ are $ω$-cocontinuous.
} then they assemble into a left adjoint to the forgetful functor
$U : Σ\trm{-monoid} → C$.
Such a theorem implies an initiality theorem since left adjoints preserve initial
objects, hence the free $Σ$-monoid over the initial object is an initial $Σ$-monoid.
Every binding signature having an initial model for $[\mb{F},\Set]$ is then
discussed as an instance of this result.
$Σ$-monoids and the theorems above have been extended by Fiore and
Saville in \cite{ListObjects17} to the larger class of $T$-monoids.
$T$-monoids are very similar to $Σ$-monoids, except that $T$ is assumed to be a
strong monad, which enables one to account for several universal algebra notions.
In this work, we have no particular use for $T$-monoids in full generality,
but \cite{ListObjects17} is still interesting in two regards.
First, the work weakened the closedness conditions in the adjoint theorem to an
$ω$-cocontinuity condition.
Second, it provided more detail on how to prove the adjoint theorem
from which one can derive the initiality theorem.

We discuss the initiality theorem and the adjoint theorem further in
\cref{subsubsec:rw-co-vs-adj}.
Furthermore, we explain the difference between,
on the one hand, the proof based on
parametrized initiality and, on the other hand,
the proof presented in \cref{subsec:building_initial_model}
based on hss, in \cref{subsubsec:rw-param-vs-hss,subsubsec:rw-building-param-hss}.

\subsubsection{Cocontinuity vs.\ existence of adjoints}
\label{subsubsec:rw-co-vs-adj}

The results of \cite{FPT99,SecondOrderDep08,ListObjects17} are similar to the
one presented in \cref{sec:initiality_theorem}, up to two differences.

First, as in \cite{ListObjects17} but in contrast to prior work, we do not
require the monoidal product to be closed; instead, we require it to preserve some
colimits, which is a weaker assumption as it is implied by having an adjoint.
For the initiality theorem, we require instead precomposition $\_ ⊗ Z$ to
preserve initial objects, binary products and $ω$-colimits.
For the adjoint theorem, we additionally require for postcomposition $Z ⊗ \_$ to
preserves $ω$-colimits.
This is an essential step towards relating the different approaches, since, in contrast to
$[F,\Set]$, precomposition does not have a right adjoint on $[\Set,\Set]$.
A result based on the existence of adjoints would hence fail to be instantiated to $[\Set,\Set]$ even
though it is a basic instantiation of the literature on modules over monads and
heterogeneous substitution systems.

\begin{proposition}
  Precomposition $\_ ∘ Z$ on $[\Set,\Set]$ does not have a right adjoint.
\end{proposition}
\begin{proof}
If precomposition by $Z$ had a right adjoint $R_Z : [\Set,\Set] → [\Set,\Set]$,
then $\Set → \Set$ would be locally small, since using that $\Id = y_1$ and the Yoneda lemma,
we would have: $\mrm{Hom}(Z,G) \cong \mrm{Hom}(\Id ∘ Z,G) \cong \mrm{Hom}(\Id,R_Z(G))
\cong \mrm{Hom}(y_1,R_Z  G) \cong R_Z(G)(1)$.
In which case, by a theorem by Freyd and Street \cite{FreydStreet95}, the
category $\Set$ would be essentially small, i.e equivalent to a small category,
which is impossible as the cardinals which are sets do not form a set.
\end{proof}

Second, the authors proved the adjoint theorem (c.f.\ \cref{thm:adjoint-theorem})
and deduced the initiality theorem (c.f.\ \cref{thm:initiality-theorem})
from it, using that left adjoints preserve initial objects.
We have done the opposite: we proved the initiality theorem, and deduced the
adjoint theorem from it (c.f.\ \cref{subsec:building-adjoint}).
This enables us to prove the initiality theorem using the underlying functor
$I + H\_$ rather than the functor $I + H\_ + X ⊗ \_$, which enables us to
remove the hypothesis that $X ⊗ \_$ is $ω$-cocontinuous in the
hypothesis of the initiality theorem.
Even though the results presented here are only about monoidal categories,
the removal of this hypothesis is important for
dealing with some skew-monoidal categories as discussed in
\cref{subsubsec:rw-skew}.

\subsubsection{Parametrized initiality vs.\ heterogeneous substitution systems}
\label{subsubsec:rw-param-vs-hss}

Since the use of parametrized initiality for $Σ$-monoids is explained in little
detail in \cite{FPT99,SecondOrderDep08,ListObjects17}, the differences of
vernacular and presentation can lead one to believe that the proof based on
parametrized initiality is fundamentally different from the one based on hss
described in \cref{sec:building_initial_model}.

However, hss and parametrized initiality are actually very similar, and the
proofs are strongly related; even though to the best of our knowledge this has
never been reported before.
To understand the differences, let us consider the proof of the initiality
theorem, starting from parametrized initiality.

\begin{definition}[Parametrized Initiality]
  Let $F : \mc{C} × \mc{C} → \mc{C}$ be a bifunctor with pointed
  strength $\mrm{st}$, and $U : \mc{C}$.
  An $F(U,\_)$-algebra $(R,r)$ is \emph{parametrized initial} if for
  any pointed object $Z : \mc{C}$ and $F(U ⊗ Z,\_)$-algebra $(C,c)$,
  there is a unique morphism $h : R ⊗ Z → C$ such that the following diagram commutes.
  \[
    \begin{tikzcd}[column sep=large]
      F(U, R) ⊗ Z \ar[r, "\mrm{st}"] \ar[d, swap, "r ⊗ Z"]
        & F(U ⊗ Z, R ⊗ Z) \ar[r, dashed, "F(U ⊗ Z{{,}} h)"]
        & F(U ⊗ Z, C) \ar[d, "c"] \\
      R ⊗ Z \ar[rr, swap, dashed, "h"]
        &
        & C
    \end{tikzcd}
  \]
\end{definition}

\noindent To prove the initiality theorem, we are interested in parametrized
initiality for $U := I$, and for the bifunctors of the form $F : (U,A)
\longmapsto U + H(A)$ such that $H$ is a functor with pointed strength $Θ$.
In this case, assuming that $\_ ⊗ Z$ distributes over binary products,
parametrized initiality unfolds to \hyperref[def:hss]{hss}, except that for
parametrized initiality, the output of $h$ can be any $H$-algebra $C$ with a map
$f : Z → C$, whereas it is fixed to be $R$ for hss.
Therefore, following the proof of \cref{subsec:hss_models}, both
parametrized initiality and hss induce models.
\[
  \begin{tikzcd}[column sep=large]
    I ⊗ Z \ar[r, "η ⊗ Z"] \ar[dd, swap, "\lambda_Z"]
      & R ⊗ Z \ar[dd, dashed, "h"]
      & H(R) ⊗ Z \ar[l, swap, "r ⊗ Z"]
                       \ar[d, "Θ_{R,e}"] \\
      &
      & H(R ⊗ Z) \ar[d, dashed, "H(h)"] \\
    Z \ar[r , swap, "f"]
      & C
      & H(C) \ar[l, "r"]
  \end{tikzcd}
\]

The main difference lies in the initiality part of the proof, more specifically
in the proof that the initial algebra morphism respects the monoid multiplication.
For parametrized initiality, this can be directly proven by appropriately
instantiating $Z$ and $C$.
However, as $C$ is fixed to be $T$ in the definition of hss, it forces us
to use a fusion law as done in \cref{subsec:building_initial_model}.

This makes no difference for wellfounded syntax, but it is fundamental when
it comes to non-wellfounded syntax.
Indeed, as parametrized initiality automatically yields an initial model, it can
not be used to prove that non-wellfounded languages have a monadic substitution
structure; this is in contrast to hss which were designed to handle both
wellfounded and non-wellfounded syntax \cite{Hss04,HssNonWellfounded24}.
Consequently, to better relate all the different approaches, we have decided to
base our work on hss in \cref{sec:building_initial_model}.

\subsubsection{Building parametrized initial algebras and heterogenous substitution systems}
\label{subsubsec:rw-building-param-hss}

It remains to understand how hss and parametrized initiality are built out of initial algebras.
As explained in \cref{subsec:building_hss} generalizing \cite{HssUntypedUniMath19},
hss can be built out of initial algebras using generalized Mendler's style
iteration (\cref{thm:gen-mendler}), by instantiating them with $F := I + H\_$,
$L := \_ ∘ Z$, $X := R$ and an appropriate $Ψ$.
Rather than using Mendler's style iterations, Fiore and Saville defined instead
a ``lax-uniformity property of initial algebra functors'' in \cite[Theorem 4.7]{ListObjects17}.
We depart from their presentation, as we noticed that the hypothesis and
conclusions could actually be separated in a theorem and a corollary.

\begin{theorem}
  \label{thm:FioreSaville}
  Let $\mc{A,B,C,D}$ be categories such that $\mc{C}$ has an initial object and $ω$-colimits.
  Let $F,G,K,J$ be functors and $t : J ∘ F → G ∘ (K × J)$ a natural
  transformation as below left, such that:
  \begin{itemize}[label=$-$]
    \setlength\itemsep{-1pt}
    \item for all $D : \mc{D}$, $F(D,\_)$ is $ω$-cocontinuous with initial algebra $(μF(D),r_{μF})$
    \item $J$ preserves initiality and $ω$-colimits
  \end{itemize}
  Then, for any $G(K(D),\_)$-algebra $(C,c)$ there exists a unique morphism
  $h : J(μF(D)) → C$ such that the diagram below right commutes:
  \begin{align*}
    \begin{tikzcd}[ampersand replacement=\&]
      \mc{D} × \mc{C} \ar[r, "F"] \ar[d, swap, "K × J"]
        \& \mc{C} \ar[d, "J"] \ar[dl, Rightarrow, shorten=13pt, swap, "t"] \\
      \mc{B} × \mc{A} \ar[r, swap, "G"]
        \& \mc{A}
    \end{tikzcd}
    &&
    \begin{tikzcd}[ampersand replacement=\&, column sep=large]
      J(F(D,μF(D))) \ar[r, "t"] \ar[d, swap, "J(r_D)"]
        \& G(K(D),J(μF(D))) \ar[r, dashed, "G(K(D){,} h)"]
        \& G(K(D),C) \ar[d, "c"]\\
      J(μF(D)) \ar[rr, swap, dashed, "h"]
        \&
        \& C
    \end{tikzcd}
  \end{align*}
\end{theorem}

\begin{corollary}
  Suppose additionally that $\mc{A}$ has an initial object and $ω$-colimits, and
  that for all $B : \mc{B}$, $G(B,\_)$ is $ω$-cocontinuous with initial algebra $(μG(B),r_{μG})$.
  Then, $μF$ and $μG$ assemble into functors $μF : \mc{D} → \mc{C}$ and $μG : \mc{B} → \mc{A}$,
  and there is a natural transformation $h : J ∘ μF → μG ∘ K$ defined pointwise as the unique morphism
  associated to the initial algebra $μG_{K(D)}$:
  \begin{align*}
    \begin{tikzcd}[ampersand replacement=\&]
      \mc{D} \ar[r, "R"] \ar[d, swap, "K"]
        \& \mc{C} \ar[d, "J"] \ar[dl, Rightarrow, shorten=13pt, swap, "h"] \\
      \mc{B} \ar[r, swap, "B"]
        \& \mc{A}
    \end{tikzcd}
    &&
    \begin{tikzcd}[ampersand replacement=\&, column sep=large]
      J(F(D,μF(D))) \ar[r, "t"] \ar[d, swap, "J(r_{μF})"]
        \& G(K(D),J(μF(D))) \ar[r, dashed, "G(K(D){,} h)"]
        \& G(K(D),μG_{K(D)}) \ar[d, "r_{μG}"]\\
      J(μF(D)) \ar[rr, swap, dashed, "h_D"]
        \&
        \& μG_{K(D)}
    \end{tikzcd}
  \end{align*}
\end{corollary}

Doing so is actually fundamental to relate the approaches.
Indeed, as noticed in \cite{CoqPl2023MonCatHss}, the extra assumptions of the
corollary $\mc{A}$ and $G$ are not needed to prove the initiality theorem.
Now that these hypotheses are separated in a corollary, we can go even further.
Though it might be surprising at first sight due to differences of presentation,
Mendler's style iterations and \cref{thm:FioreSaville} are actually direct
instantiations of each other, and hence simply different formulations of the
same theorem.
Consequently, the differences between the two proofs fully boil down to the
differences between using hss or parametrized initiality as discussed in
\cref{subsubsec:rw-param-vs-hss}.

\begin{theorem}
  \cref{thm:FioreSaville} is an equivalent formulation of Mendler's style iteration (\cref{thm:gen-mendler}).
\end{theorem}
\begin{proof}
  As remarked in \cite{CoqPl2023MonCatHss}, \cref{thm:FioreSaville} can be deduced
  from Mendler's style iteration by setting $F := F(D,\_)$, $L := J$, $X := C$ and
  $Ψ\;h \mapsto c ∘ G(KD,C) ∘ t$.

  Conversely, as noticed by Lafont, Mendler's style iteration can be deduced
  from the \cref{thm:FioreSaville} by setting $\mc{B,D} := 1$, $\mc{C} :=
  \mc{C}$, $\mc{A} := \Set^\op$, and $F := F$, $L := \mc{D}(L(\_),X)$.
  This enables us to set $t := Ψ$ by seeing $Ψ$ as a natural transformation of
  type $C → \Set^\op$ rather than of type $C^\op → \Set$.
  Then, applying \cref{thm:FioreSaville} to the trivial algebra $1 : \Set^\op$
  gives us a diagram in $\Set^\op$, or in $\Set$ as right-below, which provides
  us with Mendler's style iteration when evaluated in $\star : 1$.
  \begin{align*}
    \begin{tikzcd}[ampersand replacement=\&]
      \mc{C} \ar[r, "F"] \ar[d, swap, "\mc{D}(L(\_){,}X)"]
        \& \mc{C} \ar[d, "\mc{D}(L(\_){,}X)"] \ar[dl, Rightarrow, shorten=13pt, swap, "Ψ"] \\
      \Set^\op \ar[r, swap, "\Id"]
        \& \Set^\op
    \end{tikzcd}
    &&
    \begin{tikzcd}[ampersand replacement=\&]
      \mc{D}(L(F(R)),X)
        \& \mc{D}(L(R),X)  \ar[l, swap, "Ψ"]
        \& 1 \ar[l, swap, dashed, "h"] \\
      \mc{D}(L(R),X) \ar[u, "L(r)^*"]
        \&
        \& 1 \ar[u] \ar[ll, dashed, "h"]
    \end{tikzcd}
  \end{align*}
\end{proof}

\subsubsection{Applications}

As the framework based on $Σ$-monoids is defined, from the outset, for monoidal
categories, it can be, and has been, applied to more involved instances than
$[\mb{F},\Set]$ and untyped languages.
However, note that as until \cite{ListObjects17} this framework relied on
closedness, most applications have been restricted to closed monoidal categories.
This excludes instances like $[\Set,\Set]$, that were hence replaced by
categories like $[\mb{F},\Set]$ that are closed.

Simply-typed languages where first investigated in \cite[Section II.1.1]{Cbn02},
where Fiore explained that the simply-typed lambda calculus is an initial algebra
on $[\mb{F} ↓ T,\Set]^T$, and suggested that substitution could be accounted for
using the framework of \cite{FPT99}.
This was briefly detailed in \cite[Section 1.3]{MMCCS05}, where Fiore
additionally discussed the monoidal structure on $[\mb{F} ↓ T,\Set]^T$.
It was worked out fully and in great detail by Zsidó in her dissertation
\cite[Chapter 5]{ZsidoPhd10}.
She fully proved, without using any high-level theorem, that simply-typed
algebraic signatures induce strong endofunctors on $[\mb{F} ↓T,\Set]^T$, that
the category $[\mb{F} ↓T,\Set]^T$ is a left-closed monoidal category, and that
these signatures induce initial $Σ$-monoids in that category.

Hamana investigated polymorphic languages in \cite{Polymorphism11}.
For System $F$ and System $F_ω$, Hamana built categories $∫ G$ and $∫ H$ such that
the System $F$ and System $F_ω$ are initial algebras on $\Set^{∫G}$ and $\Set^{∫H}$, respectively.
He then introduced polymorphic and higher-order polymorphic counterparts to
algebraic signatures, and generalized the constructions to them.

Fiore also suggested an initial framework for dependently typed languages in
\cite[Section II]{SecondOrderDep08}.

\subsubsection{Extension to skew-monoidal categories}
\label{subsubsec:rw-skew}

Not all categories relevant to initial algebra semantics are monoidal.
A category to interest is $ℕ$-indexed families of sets \cite{CellularHoweTheorem20}, i.e. functors $[\N,\Set]$,
with the tensor product $A(n) ⊗ B(n) := ∑_{m : ℕ} A(m) × B(n)^m$.
Yet, this tensor is not monoidal but only skew-monoidal as the associator and
the unit are not isormorphisms.
Therefore, to apply Fiore's framework of $Σ$-monoids on $[\N,\Set]$
Borthelle, Tom Hirschowitz\footnote{
  Be aware that both André Hirschowitz and Tom Hirschowitz worked on initial
  semantics, sometimes in joint work. To distinguish them, we refer to them by
  their full name.
} and Lafont generalized it to skew-monoidal categories \cite{CellularHoweTheorem20},
and formalized the results in Coq.

Modules over monoids seem straigthfoward to generalize skew-monoidal categories;
however, as of yet, the proofs based on hss and parametrized initiality do not
seem to generalize.
Indeed, we can no longer prove the monoid's associativity law that is an
equality of morphisms of type $(R ⊗ R) ⊗ R → R$, by instantiating hss
and parametrized initiality for $Z := R ⊗ R$.
Basically, the reason is that this instantiation provides uniqueness of a
morphism of type $R ⊗ (R ⊗ R) → R$, but as the associator $α$ is not invertible
in skew-monoidal categories, this is no longer equivalent to uniqueness of a
morphism of type $(R ⊗ R) ⊗ R → R$.
Thus, we could have generalized the framework presented here by adapting the
proof of \cite{CellularHoweTheorem20} that repeatedly applies \cref{thm:FioreSaville},
but that would have been at the cost of not unifying hss with the other frameworks.
As our goal is to unify the different approaches, we have chosen not to do so, and
leave open the challenge to generalize hss and this framework to skew-monoidal categories.

Skew-monoidal categories were also considered in \cite{NamelessDummies22} by
André and Tom Hirschowitz, Lafont, and Maggesi.
Therein, they study De Bruijn monads and De Bruijn S-algebras, for untyped and
simply-typed languages, and identify them as monoids and $Σ$-monoids in the
skew-monoidal category $[1,\Set]$ for the relative functor $J : * ↦ ℕ$,
and $[1,\Set^T]$ for the the relative functor $J : * ↦ t ↦ \N$.

Unfortunately, the initiality theorem proven in \cite{CellularHoweTheorem20}
for skew-monoidal categories fails to apply to their instances, as it ``requires
that the tensor product is finitary in the second argument'' which is not the
case for their instances.
Consequently, they had to reprove an initiality theorem from scratch specifically for their case.
This issue actually arose because they proved an adjoint theorem, and
deduced an initiality theorem from it, rather than doing the opposite.
As discussed in \cref{subsubsec:rw-co-vs-adj}, if they had done the opposite,
they could have removed the additional hypothesis, and reused the theorem.

\subsubsection{Further work on $Σ$-monoids}

Though this is not the subject of this article, works on $Σ$-monoids
have been developed further to account for meta-variables and equations.
An account of meta-variables for $Σ$-monoids was first developed by Hamana in
\cite{HamanaMetavar04}, and by Fiore in \cite[Section I.2]{SecondOrderDep08}.
Equational systems were then considered in Hur's dissertation \cite{HurPhd}
supervised by Fiore, which was reported in \cite{FioreHur07,FioreHur09,FioreHur08},
and partially extended in Hur's dissertation.
Equations and second-order languages were also considered in
\cite{FioreHur10,FioreMahmoud10}, and for polymorphic languages in
\cite{PolymorphismEq13}.
This work was recently used to design an Agda framework for reasoning with
higher-order languages and substitution in \cite{FioreSzamozvancevPopl22}.

\subsection{Modules over Monads}
\label{subsec:rw-modules-over-monoids}

\subsubsection{Origins}

Modules over monads were studied by André Hirschowitz and Maggesi in
\cite{HirschowitzMaggesi07}, in order to capture the substitution structure of
higher-order languages and their constructors.

\begin{definition}[Module over a monad]
  Given a monad $(R,η,μ) : \Mon([\mc{C},\mc{C}])$ on $\mc{C}$,
  a module over $R$ with codomain $\mc{D}$ is a tuple $(M,p^M)$ consisting
  of a functor $M : \mc{C} → \mc{D}$ and a natural transformation $p : M ∘ R → M$
  compatible with $η$ and $μ$ as in \cref{def:modules}.
\end{definition}
The authors consider monads on $\Set$ and modules with codomain $\Set$
to define models of a given signature as pairs of a monad $R$ together with a
family of suitable module morphisms over $R$,
and stated that every untyped algebraic signature has an initial model.
They also considered how modules over monads could potentially encompass more
notions, such as simply-typed syntax, or some form of semantic properties.
This work was refined in \cite{HirschowitzMaggesi10}, where it has been
enriched with a proof, based on syntax trees, that is specific to
$[\Set,\Set]$ and to untyped algebraic signatures.

Zsidó, in her Ph.D.~\cite[Chapter 6]{ZsidoPhd10}, extended this approach
to simply-typed languages on $[\Set^T,\Set^T]$, studying modules with codomain $\Set$.
First, she showed that modules over monads do encompass simply-typed
algebraic signatures and extended the notion of model to support them.
Second, she extended the proof based on syntax trees to prove that every
simply-typed algebraic signature has an initial model.
This result was formalized in the proof assistant Coq by Ahrens and Zsidó \cite{ISCoq10}
using Coq's built-in inductive types to represent languages.

\subsubsection{Modules over Monads vs Modules over Monoids}
\label{subsubsec:module_monads_vs_monoids}

Compared to the original work on the subject \cite{HirschowitzMaggesi10,ZsidoPhd10},
our framework does not rely on modules over monads but on modules over monoids
(\cref{def:modules}) in a monoidal category.
This enables us to account for categories like $[\mb{F},\Set]$, whose objects are not endofunctors,
whereas a framework modules over monads are limited to endofunctor categories.
Indeed, for a monoidal category as $\_ ⊗ \_ : \mc{C} × \mc{C} → \mc{C}$, $M ⊗ R$
is only well-defined for $M : \mc{C}$, and thus $M$ and $R$ can not be of
different types.
Moreover, even for functor categories $[\mc{B},\mc{C}]$, it is not definable as
$M ∘ R : \mc{B} → \mc{D}$ and $M : \mc{C} → \mc{D}$ are of different types, so
there can not be a natural transformation from $M ∘ R$ to $M$.

In the case of endofunctor categories $[\mc{C},\mc{C}]$, modules over monads
are more general than modules over monoids, as modules over monoids are
exactly modules over monads with codomain $\mc{C}$.
The added power of choosing $\mc{D}$ was thought to handle the simply-typed
case, and indeed, in \cite[Chapter 6]{ZsidoPhd10} modules over monads on
$[\Set^T,\Set^T]$ with codomain $\Set$ are used to model simply-typed algebraic
signatures.
The idea here is that the output set of modules can be used to represent the
\emph{terms of a specific type} of the inputs and output of the constructors.
Nevertheless, in practice modules over monads do not seem to add any
expressive power.
In the untyped case, for the category $[\Set,\Set]$
\cite{HirschowitzMaggesi07,HirschowitzMaggesi12,PresentableSignatures21,ReductionMonads20},
$\mc{D}$ is always chosen to be $\Set$, in which case both notions
coincides.
In the simply-typed case, as will be explained in \cref{subsubsec:rw-gen-strength},
modules over monoids are enough to represent the languages that have been considered, and the added
flexibility of modules over monads is not necessary.
Moreover, as explained in \cite[Section 2.4]{TransitionMonads22}, for any
type $t : T$ there is an adjunction on the functor categories that yields
an adjunction between modules over monads and modules over monoids:
\begin{align*}
  \begin{tikzcd}[ampersand replacement = \&, column sep=large]
    [\Set^T{,}\Set]            \ar[r, bend left, "y(t) ∘ \_"]
                               \ar[r, phantom, "\perp"]
      \& {[}\Set^T{,}\Set^T{]} \ar[l, bend left, "(\_)_t ∘ \_"]
  \end{tikzcd}
  &&
  \begin{tikzcd}[ampersand replacement = \&, column sep=large]
    \Mod(\Set^T{,}\Set)
        \ar[r, bend left]
        \ar[r, phantom, "\perp"]
      \& \Mod(\Set^T{,}\Set^T) \ar[l, bend left]
  \end{tikzcd}
\end{align*}
Here $y(t) : T → \Set^T$ is the Yoneda embedding, and $(\_)_t : \Set^T → T$ is
the projection.
This adjunction implies that modelling an algebraic constructor using modules
over monads with codomain $\Set$ as done in \cite[Chapter 6]{ZsidoPhd10} is the
same as using modules over monoids.
Indeed, the monoid underlying the model is the same in both representations,
and the choice of module morphism corresponds under the adjunction.
Therefore, in practice, for simply-typed languages there is actually no difference
in terms of models between using modules over monads or modules over monoids.

However, using modules over monoids makes a real difference when it comes to
building an initial model.
In the case of modules over monoids, using that all the outputs of the
constructors can be set to the trivial signature $Θ$, we summed the
inputs into one single signature $Σ$, such that we can represent our
constructors as a single morphism of modules $Σ → Θ$.

This is required to apply the proof of \cref{sec:building_initial_model} based
on hss, as it requires an initial algebra for the functor $\ul{I} + H$.
It is not possible for modules over monads, as already for algebraic signatures
the modules describing the outputs $[Θ]_t$ depend on the output types $t$, and
hence are all different.
Therefore, it is not possible to directly apply the proof of
\cref{sec:building_initial_model}.
This is important as the proof using syntax trees is less modular and
significantly longer than the one using hss, and is specific to $[\Set^T,\Set^T]$
whereas the proof using hss applies to any appropriate monoidal category.

Last but not least, modules over monads have the advantage of directly connecting
to the work of Fiore and hss as described in \cref{related-work:model-sigma-monoids}.
For all theses reasons, we prefer modules over monoids to modules over monads.

\subsubsection{Earlier attempt at comparing traditions: Zsidó's Ph.D.\ thesis}

Though little-known, one of the most important work for our purpose of
relating the different traditions is Zsidó's dissertation \cite{ZsidoPhd10}.
There, Zsidó relates Fiore's work \cite{FPT99,Cbn02} to Hirschowitz and
Maggesi's work \cite{HirschowitzMaggesi07,HirschowitzMaggesi10} on initial
semantics; these use $[\mb{F},\Set]$ and $[\Set,\Set]$, respectively, in the
untyped case, and $[\mb{F}↓T]^T$ and $[\Set^T,\Set^T]$, respectively, in the simply-typed case.

In the untyped case, where modules over monads coincide with modules over
monoids, Zsidó generalises modules over monoids and models from $[\Set,\Set]$
to monoidal categories, much like in \cref{subsec:modules,subsec:models}, in
order to express the work on $[\mb{F},\Set]$ and $[\Set,\Set]$ in the same
language.
She then proves, by appropriately propagating the adjunction between
$[\mb{F},\Set]$ and $[\Set,\Set]$ throughout the framework, that for any
algebraic signature, there is an initial model on $[\mb{F},\Set]$ if and
only if there is one for $[\Set,\Set]$, and that they can be constructed
from one another.

In the simply-type case, she starts by reviewing the work of Fiore
\cite{Cbn02} in \cite[Chapter 5]{ZsidoPhd10}, and by completing
the work of André Hirschowitz and Maggesi sketched in
\cite[Section 6.3]{HirschowitzMaggesi07} in \cite[Chapter 6]{ZsidoPhd10}.
However, in her work on simply-typed syntax, Zsidó uses modules over monads
with codomain $\Set$, preventing her from defining both frameworks in the same
language, and fully relating the two approaches.
Consequently, in \cite[Chapter 7]{ZsidoPhd10}, she only provides
an adjunction between $Σ$-monoids and models.
Furthermore, the adjunction on models no longer comes from parametricity in
the input monoidal category but is now lifted in an ad-hoc way.

\subsubsection{Signatures}
\label{subsubsec:rw-modules-sig}

In all previous work we have discussed on modules over monads, all the signatures
considered were purely syntactic, either untyped algebraic signatures
or simply typed algebraic signatures.
The general notion of signatures (\cref{def:sig}) was introduced by André
Hirschowitz and Maggesi in \cite{HirschowitzMaggesi12} on $[\Set,\Set]$.
The authors further provided their signatures with the modularity property that
pushout of signatures lift to initial models (\cref{prop:modularity-models}), and
established that signatures with strength induce signatures similarly to
\cref{prop:sigstrength_to_sig}.

This work was expanded on and formalized in Coq in \cite{PresentableSignatures21},
still on $[\Set,\Set]$, by Ahrens, André Hirschowitz, Lafont and Maggesi.
Moreover, they introduced \emph{presentable} signatures: a signature $Σ$ is
presentable iff there exists an algebraic signature $Υ$ with an epimorphism $Υ →
Σ$. The authors prove that presentable signatures are representable.
Compared to algebraic signatures, presentable signatures
allow us to model constructors with \emph{semantic} properties, such as a binary
\emph{commutative} operator.
In particular, they enable us to do this without the need for any further construction in the framework.
In contrast, 2-signatures \cite{2Signatures19} introduce an explicit notion of equation between terms.
In practice, examples of presentable signatures are covered by signatures
with strength as the latter are stable under some colimits (\cref{prop:sigstrength_omega_cocomplete}),
but a more precise link is still to be established.

In the two previously cited papers \cite{PresentableSignatures21,2Signatures19}, the authors worked on $[\Set,\Set]$ where
modules over monoids and modules over monads coincide, and where constructors
can be represented by morphisms $Σ → Θ$.
Since this is not possible for modules over monads, many authors
\cite{ZsidoPhd10,ISCoq10,ExtendedInitiality12,UntypedRelativeMonads16,TypedRelativeMonads19}
only considered syntactic signatures, and defined a model to be not a monoid
with a single module morphism, but a monoid with a family of module morphisms.
As we explain in \cref{subsubsec:rw-gen-strength}, this is not necessary for our
purpose, and not doing so enables us to consider more general notions of signatures.

More generally, it would also be possible to require a second signature to
specify the output rather than using $Θ$.
However, as we have no use for non trivial outputs, and as it simplifies the
framework, we limit ourselves to $Θ$ for the output signature.

\subsubsection{A generalized recursion principle for simply-typed languages}
\label{subsubsec:rw-gen-rec}

As explained in \cref{subsubsec:rep-sig-rec}, we care about \emph{representable} signatures,
i.e., signatures that admit an initial model, as this provides languages with
a recursion principle.
In the untyped case this works well as exemplified in \cref{ex:FOL-to-LL}, where we use
the recursion principle to build a substitution-safe translation from first-order logic to linear logic.
However, for it to work we implictly use that we can build a model of first-order
logic out of the model of linear logic, as they have both have a model in
the \emph{same} monoidal category $[\Set,\Set]$.
In the simply-typed case, doing so is very limited, since the type system $T$ is
hardcoded in the monoidal category, as in $[\Set^T,\Set^T]$.
Consequently, two different languages with different type systems $T$ and $T'$
have models in different monoidal categories $[\Set^T,\Set^T]$ and $[\Set^{T'},\Set^{T'}]$,
and hence can not be directly related by initiality as in the untyped case.
In a sense, this is not very surprising as such a translation should first rely
on a translation $T → T'$ of the type system which is not part of our framework.

To equip simply-typed languages with better recursion principles, in a
technical paper \cite{ExtendedInitiality12}, Ahrens reworked the entire
framework to internalize the typing system $T$ in the framework.
He replaced monads by ``$T$-monads'', adapted modules, signatures, and models,
before proving an initiality theorem.
He then formalized this framework in Coq and used it to provide a verified
translation of PCF into the untyped lambda calculus.
While this framework is interesting, it is very specific to endofunctor
categories $[\Set^T,\Set^T]$ and it is unclear how it would scale to more
involved languages and how it relates to the framework presented here.

\subsubsection{Further work on initial semantics}

Though this is not the subject of this article, modules over monads have
been used further to add semantics in the form of reduction rules between terms on top of the abstract-syntax framework
that is described in this work.
Modules over relative monads and a notion of 2-signature were used in
\cite{UntypedRelativeMonads16,TypedRelativeMonads19}, to model context-passing reduction
rules for untyped and simply-typed algebraic signatures.
Similar notions were used in \cite{2Signatures19} to add equations in the
untyped case.
The original work on reduction rules by Ahrens was extended in the untyped
case to handle conditional rules as top-level $\beta$-reduction
\cite{ReductionMonads20},
and later extended to account for simply-typed languages and reduction strategies, e.g.,
call-by-value reduction in \cite{TransitionMonads20,TransitionMonads22}.

\subsection{Heterogeneous Substitution Systems}
\label{subsec:rw-hss}

\subsubsection{Origins}

Heterogeneous substitution systems (\cref{def:hss}) were introduced on
endofunctor categories $[\mc{C},\mc{C}]$ by Matthes and Uustalu in \cite{Hss04}.
Studying pointed strong functors as signatures, they design hss as an intermediate
abstraction to prove that both wellfounded or non-wellfounded higher-order
languages have a well-behaved substitution structure, that is, form monads
\cite[Theorem 10]{Hss04}.
In the well-founded case, they prove that assuming that precomposition $\_ ∘ Z$ has a
right-adjoint for all $Z$, then any signature with strength that has an initial
algebra has an associated hss, and as such a monad structure \cite[Theorem 15]{Hss04}.
In the non-wellfounded case, they have proved then any signature with
strength that has a cofinal algebra has an associated hss, and as such a
monad structure \cite[Theorem 17]{Hss04}.

Still on $[\mc{C},\mc{C}]$, this work was strengthened into a framework for
initial semantics by Ahrens and Matthes in \cite{HssRevisited15}.
They assemble signatures with strength and hss in categories, prove a fusion law
(\cref{thm:fusion-law}) for generalized Mendler's style iteration \cite[Lemma10]{HssRevisited15},
and use it to prove that the hss built in \cite{Hss04} is actually initial as an hss.
This yields a framework for initial semantics based on hss, which they
additionally formalized Coq using the UniMath library \cite{UniMath}.
Compared to them, we have chosen to keep hss as an intermediate abstraction
in the proof rather than making it our notion of model.
First, it enables us to have a common notion of model connecting the work
of Fiore, modules over monads, and hss.
Second, hss provide less explicit information on substitution than
monoids, and, importantly, have a stronger recursion
principle that is applicable in fewer situations, and that we have found no specific use for.

Hss were applied in \cite{HssUntypedUniMath19} by Ahrens, Matthes, and Mörtberg
to build untyped higher-order languages from elementary type constructors in the
UniMath library, and to prove that they have a monadic substitution structure.
The proof of \cite[Theorem 10]{Hss04} is based on left-closedness of the monoidal structure and the
associated version of Mendler's style iterations \cite[Theorem 2]{GeneralisedFold99}.
Yet, as discussed in \cref{subsubsec:rw-co-vs-adj}, $[\Set,\Set]$ is not left-closed.
Consequently, to build an hss structure, and hence a monad structure, the
authors turned to $ω$-cocontinuity and the associated version of Mendler's style
iterations \cite[Theorem 1]{GeneralisedFold99}.
However, they did not prove that the built hss is initial.

The proof presented in \cref{sec:building_initial_model} is a generalization of
the proof of \cite{HssUntypedUniMath19} to monoidal categories for the construction
of the (initial) hss and hence of the monad structure.
To prove that the resulting model is initial, we generalize the proofs of \cite{Hss04,HssRevisited15}
to monoidal categories and $ω$-cocontinuity, and adapt them from hss to models.
Doing so, we provide more detailed proofs then currently available in the litterature.

\subsubsection{Generalized strength}
\label{subsubsec:rw-gen-strength}

Hss were also applied on $[\Set^T,\Set^T]$ by Ahrens, Matthes and Mörtberg
in \cite{HssTypedUnimath22} to build simply-typed higher-order languages in the
UniMath library \cite{UniMath}, akin to \cite{HssUntypedUniMath19}.
To be able to build simply-typed signatures modularly, they introduced the
concept of \emph{generalized strength} \cite[Definition 2.15]{HssNonWellfounded24}:
\begin{definition}[Generalized Strength]
  A \emph{generalized strength} for a functor $H : [\mc{C},\mc{D}'] → [\mc{C},\mc{D}]$
  is a natural transformation $Θ$ such that for all $A : [\mc{C},\mc{D}']$,
  and pointed object $b : \Id → B$ with $B : [\mc{C},\mc{C}]$,
  $Θ_{A,b}$ has the type
  \[ Θ_{A,b} : H(A) ∘ B \longrightarrow H(A ∘ B) \]
  and satisfies associativity and unit laws as in \cref{def:sig-strength}.
\end{definition}
Similarly to signatures with strengths (\cref{subsec:sigstrength-to-sig}),
generalized signatures with strength correspond correspond to modules over
monads when $\mc{D}' := \mc{C}$.
\begin{proposition}
  Given an endofunctor $H : [\mc{C},\mc{C}] → [\mc{C},\mc{D}]$ with generalized
  strength $Θ$, for any monad $(R,η,μ)$, $H(R)$ is a module over the monad $(R,η,μ)$
  with codomain $\mc{D}$ with the action $H(R) ∘ R \xrightarrow{Θ_{R,η}} H(R ∘
  R) \xrightarrow{H(μ)} H(R)$
\end{proposition}

\noindent Nonetheless, their work does not amount to working with modules over
monads as they only use generalized strength as an intermediate tool to build
regular signatures with strength, on which they solely rely in the end.

Actually, the slightly more general notion of generalized strength, like the one of modules
over monads, is not needed to model and build simply-typed signatures modularly.
As an example, let us consider the abstraction of the simply-typed lambda calculus.
Given a term of type $t$ with a free variable of type $s$, it should return a term
of type $s → t$, i.e., $\abs_{s,t} : Λ_T (Γ + y(s))(t) → Λ_T (Γ)(s → t)$,
where $y(s)(u) = {*}$ if $u = s$ and $∅$ otherwise.
As constructors are represented by morphisms of the form $X → Θ$, we must ensure
that all fibers but $Λ_T(Γ)(s → t)$ are empty, and that in the fiber over $s → t$,
we have $X(Γ)(s → t) := Λ_T (Γ + y(s))(t)$.
We can do so by postcomposing $Θ$ with a signature with strength $δ$ that trivialize fibers,
$\swap$ to swap them, and $y(s)$ to add free variables.

  \begin{definition}
    The functor on $[\Set^T,\Set^T]$, $δ$ and $\swap$ are $ω$-cocontinuous:
    \[ \begin{array}{ccc}
      δ_s\; X\; Γ\; u \;
      = \left\{
        \begin{array}{ll}
          X(Γ)(s) & \mrm{if}\; u := s \\
          ∅       & \mrm{otherwise}
        \end{array} \right. & &

        \swap_s^t\; X\; \Gamma\; u := \left\{
        \begin{array}{ll}
          X(\Gamma)(s) & \mrm{if}\; u := t \\
          X(\Gamma)(t) & \mrm{if}\; u := s \\
          X(\Gamma)(u) & \mrm{otherwise}
        \end{array} \right.
    \end{array} \]
  \end{definition}

  \noindent We can then build an $ω$-cocontinuous signature with strength
  representing variable binding, by iterating $y(s)$ and using that
  postcomposing by, and $ω$-cocontinuous signatures, preserve $ω$-cocontinuity
  (\cref{ex:left-comp}):

  \begin{definition}
    There is a signature with strength $[Θ^l_s]_t$ representing the binding
    of $l$ variables in an input of type $s$ and returning a term of type $t$ by
    $\swap_{s → t}^t ∘ δ_{s → t} ∘ y(l) ∘ Θ$
  \end{definition}

  \noindent Using the usual closure under coproducts and finite products, we can
  then represent the simply-typed lambda calculus as follows:

  \begin{example}
    The simply-typed lambda calculus can be modeled by the $ω$-cocontinuous
    signature with stregth:
    \[ \bigplus_{s,t:T_{Λ_T}}\;\; [Θ_{s → t}]_t × [Θ_{s}]_t \; + \; [Θ^{s}_{t}]_{s → t} \]
  \end{example}

\subsubsection{Hss and monoidal categories}
\label{subsubsec:rw-hss-monoidal}

Hss have been also been applied to build the substitution structure of \emph{coinductive}
simply-typed higher-order languages by Matthes, Wullaert, and Ahrens in
\cite{HssNonWellfounded24}, in the UniMath library.
Trying to do so, Matthes faced by formalisation issues \cite{CoqPl2023MonCatHss},
in particular controlling the unfolding of definitions, e.g., of the monoidal
structure of endofunctor categories, when proving theorems.
Consequently, to be more abstract and to better control unfolding, but also to be
more general, Matthes generalized hss and the proof of the initiality
theorem to monoidal categories.
Though it is not the main subject of their paper, this was reported on in
\cite[Section 4.4]{HssNonWellfounded24}.
The proof they have formalized is basically the same as the one presented in
\cref{sec:building_initial_model} up to minor details.
Though the present papers share an author, we stress that the generalizations were
developed in parallel and independently by Matthes, and for different reasons.

In contrast to our work, \cite{HssNonWellfounded24} provides an extensive analysis of signatures
with strength in terms of actions on actegories, and uses results on actegories
as an abstract framework to equip signatures with appropriate strengths.
While this can ease formalisation, it is very technical.
Consequently, to ease understanding, we have decided to use a more direct approach.
Another difference is that their definitions are stated for the \emph{reversed}
monoidal category of ours \cite[Example 1.2.9]{2DimensionalCategories20},
i.e., for the reversed monoidal product $X ⊗' Y := Y ⊗ X$.
In practice it makes no difference as all monoidal categories are reversible,
and as they instantiate their framework with reversed categories as well.

\section{Conclusion}
\label{sec:conclusion}

In this paper, we present a framework that unifies three distinct approaches to
initial semantics -- modules over monads, signatures with strength, and
heterogeneous substitution systems -- by suitably generalizing and combining them.
Doing so, we have shown that:

\begin{enumerate}
  \setlength\itemsep{-1pt}
  \item Modules over monads provide us with an abstract and easy to manipulate framework (\cref{sec:models}).
  \item Signatures with strength enable us to state and prove an initiality theorem (\cref{sec:initiality_theorem}).
        Moreover, signatures with strength naturally appear as particular
        signatures when one tries to state such a theorem.
  \item Heterogeneous substitution systems form an intermediate abstraction that
        enables us to prove the initiality theorem modularly (\cref{sec:building_initial_model}).
\end{enumerate}

\noindent Moreover, this framework enables us to provide a detailed and
extensive discussion of related work (\cref{sec:related-work}), and to better understand the
existing literature.
Indeed, relating the different approaches and their variations to our framework
enables us to relate them to each other, clarifying a body of literature so far hard to enter.

\subsection{Open Questions}

With this work, we aim to solidify and conceptualize the foundations of
initial semantics, and to provide an accessible presentation to concepts and
proof scattered throughout many papers.
Nevertheless, there are still many open research questions in the area of
initial semantics; to consolidate the current knowledge on the subject, but
also to keep developing the framework.

\begin{itemize}
  \setlength\itemsep{-1pt}

  \item In this work, we have not considered work on metavariables and
        ``second-order'' syntax \cite{HamanaMetavar04,SecondOrderDep08}.
        Additional work is needed to properly integrate and discuss this work.

  \item This framework is formulated in terms of monoidal categories.
        Generalizing it to skew-monoidal categories would enable us to encompass
        more instances, like De Bruijn monads \cite{NamelessDummies22}.
        As discussed in \cref{subsec:rw-sigma-mon}, modules and signature with
        strength (\cref{sec:models,sec:initiality_theorem} of the framework)
        seem to generalize to skew-monoidal categories.
        However, it does not seem straightforward to generalize hss or parametrized
        initiality (\cref{sec:building_initial_model}), and work is needed to
        give a proper generalization.

  \item In this work, we have not considered semantics, only syntax.
        Yet, many frameworks have been extended to support some form of semantics.
        It remains to unify them and add them to this framework to get a full
        account of the literature on the subject:
        \begin{itemize}
          \item Many frameworks consider equations on top of syntax, like $β$-equalities.
                It remains to understand how they are related.
                For instance, it remains to understand how the approaches to equations of \cite{FioreHur10},
                and \cite{FioreSzamozvancevPopl22} relate to \cite{PresentableSignatures21,2Signatures19}.
          \item Different works have been considered to add reductions rules
                \cite{UntypedRelativeMonads16,TransitionMonads20}, and even
                reduction strategies \cite{TransitionMonads22}.
                All have an interpretation in terms of relative monads.
                Consequently, could they all be encompassed by a unique framework?
        \end{itemize}

  \item As discussed in \cref{subsubsec:rw-gen-rec}, on its own, the recursion
        principle provided by initiality is insufficient for simply-typed
        languages, as it is restricted to languages with the same type system.
        A solution was proposed in \cite{ExtendedInitiality12} by integrating
        the type system directly into the framework.
        However, this requires reworking and specializing the entire framework
        for simply-typed languages, which seems unsuitable for our purposes.
        How this issue should be solved remains an open question.

  \item Though many frameworks linked to this work have been formalized
        \cite{HssRevisited15,PresentableSignatures21,HssTypedUnimath22,FioreSzamozvancevPopl22,HssNonWellfounded24},
        this framework has not been formalized yet.
        A complete formalization should be possible and strengthen the trust in
        this framework, which can be unsettling due to its many definitions.
\end{itemize}

\section{Acknowledgements}

We thank Ambroise Lafont for fruitful discussion on initiality theorems and their different proofs.
We thank Ralph Matthes for helpful discussions on heterogeneous substitution
systems and their use to prove the initiality theorem.
We furthermore thank Marcelo Fiore for providing pointers to literature on the topic of initial semantics.
We are grateful to André Hirschowitz, Thea Li, Ambroise Lafont, and Ralph Matthes for valuable comments on early drafts.

\addcontentsline{toc}{section}{References}
\printbibliography

\end{document}